\documentclass[preprint,prc,superscriptaddress]{revtex4}
\usepackage{graphicx}
\usepackage{xcolor}
\usepackage{amsmath} 
\usepackage{amssymb} 
\usepackage[caption=false]{subfig}                                                  
\def\subfigure{\subfloat}  
\usepackage{color}          
\usepackage{booktabs}

\newcommand{\bra}[1]{\langle #1|}
\newcommand{\ket}[1]{|#1\rangle}
\newcommand{\braket}[2]{\langle #1|#2\rangle}

\begin{document}
\title{A QCD sum rules calculation of the $J/\psi D_s^* D_s$ strong coupling constant}

\author{B. Os\'orio Rodrigues}
\affiliation{Instituto de F\'{\i}sica, Universidade do Estado do Rio de
Janeiro, Rua S\~ao Francisco Xavier 524, 20550-900, Rio de Janeiro, RJ, Brazil. }
\affiliation{Centro Brasileiro de Pesquisas F\'{\i}sicas - CBPF, Rua Xavier Sigaud 150, 22290-180, Rio de Janeiro, RJ, Brazil.}

\author{M. E. Bracco}
\affiliation{Faculdade de Tecnologia, Universidade do Estado do Rio de Janeiro,
Rod. Presidente Dutra Km 298, P\'olo Industrial, 27537-000 , Resende, RJ, Brazil.}

\author{M. Chiapparini}
\affiliation{Instituto de F\'{\i}sica, Universidade do Estado do Rio de
Janeiro, Rua S\~ao Francisco Xavier 524, 20550-900, Rio de Janeiro, RJ, Brazil. }

\author{A. Cerqueira Jr.}
\affiliation{Faculdade de Tecnologia, Universidade do Estado do Rio de Janeiro,
Rod. Presidente Dutra Km 298, P\'olo Industrial, 27537-000 , Resende, RJ, Brazil.}

\begin{abstract}
In this work, we calculate the form factors and the coupling constant of the strange-charmed vertex $J/\psi D_s^* D_s$ in the framework of the QCD sum rules by studying their three-point correlation functions. All the possible off-shell cases are considered, $D_s$, $D_s^*$ and $J/\psi$, resulting in three different form factors. These form factors are extrapolated to the pole of their respective off-shell mesons, giving the same coupling constant for the process. Our final result  for the $J/\psi D_s^* D_s$ coupling constant is $g_{J/\psi D^*_s D_s} = 4.30^{+0.42}_{-0.37}\text{GeV}^{-1}$.
\end{abstract}

\maketitle

\section{Introduction}

In the past few decades, the Quantum Chromodynamics Sum Rules (QCDSR) community has been actively working in the calculation of  strong coupling constants of a great variety of vertices. Particularly, our group has  calculated a large number of charmed vertices as the $D^* D \pi$~\cite{PhysRevD.65.037502}, $D^* D \rho$~\cite{Rodrigues2011127}, $J/\psi D^* D $~\cite{doi:10.1142/S0218301305003399} among many others that were revisited in the review of ref.~\cite{Bracco20121019}. Some recent works in QCDSR target vertices with the heavy beauty mesons ($B$, $B^*$, $B_s$, $B_s^*$, ...) as the $B_s^* B K$~\cite{CerqueiraJr2012130}, $B_s^* B_s \eta^{(')}$ \cite{10.1140/epjp/i2013-13113-2}, $B_{c}^{(*)}{B}_{c}\Upsilon$ \cite{10.1140/epja/i2013-13103-y,PhysRevD.89.034017},$B_{c}^{(*)}B_{c} J/\psi$ \cite{PhysRevD.89.034017} and/or strange-charmed mesons ($D_s$, $D_s^*$, $D_{s1}$, $D_{s0}^*$,...) as the $J/\psi D_s D_s$~\cite{OsorioRodrigues2014143,PhysRevD.87.016003}, $J/\psi D_s^* D_s^*$~\cite{PhysRevD.87.016003}, $J/\psi D^*_{s0} D_{s1}$, $\phi D^*_{s0} D_{s1}$~\cite{PhysRevD.89.016001}, $D_s^* D_s \eta^{(')}$~\cite{10.1140/epjp/i2013-13113-2} and many others.

Regarding the strange-charmed vertices, which are a subject of major interest nowadays, due to the detection of various new exotic states, it is noticeable that some of them have masses that lie just above the $D_s^{(*)}D_s^{(*)}$ mass threshold. By exotic, we mean that the usual quark-model description as  $q\bar{q}$ pairs does not hold for these new states. An example of such exotic state is the $Y(4140)$, first observed by the CDF collaboration~\cite{PhysRevLett.102.242002}, which has an observed decay into the pair $J/\psi \phi$. Some interpretations for its internal structure are the tetraquark state~\cite{0954-3899-37-7-075017}, $D_s \bar{D}_s^*$ molecular state~\cite{Qiao:2013xca}, $D_s^* \bar{D}_s^*$ molecular state~\cite{Mahajan2009228,Albuquerque2009186,PhysRevD.80.017502} and hybrid state~\cite{Mahajan2009228}. Among them, we highlight the molecular $D_s^* \bar{D}_s^*$ hypothesis, so that the decay $Y(4140) \to J/\psi \phi$ can be understood with the help of an intermediate decay as $Y(4140) \to D_s^* \bar{D}_s^* \to J/\psi \phi$. In this decay, the vertex $J/\psi D^*_s D_s$ is present, thus a more precise knowledge of its coupling constant and form factors may help in the understanding of the fundamental constitution of the $Y(4140)$ meson. 
Also, it may be useful in the studies of other exotic states as the $X(4350)$~\cite{PhysRevLett.104.122001,Zhao:2011sd,0253-6102-54-6-22,PhysRevD.82.015013}, $Y(4274)$~\cite{Aaltonen:2011at} and the $Y(4660)$~\cite{PhysRevLett.99.142002,Chen:2013axa}, for example.

In this work, we investigate the $J/\psi D^*_s D_s$ vertex within the QCDSR formalism, following the development of our previous work for the $J/\psi D_s D_s$ vertex~\cite{OsorioRodrigues2014143}. This development consists in calculating the form factors from the three point correlation function in two ways: one with quarks and gluons degrees of freedom, the OPE side, and the other with hadronic degrees of freedom, the phenomenological side. After that, a double Borel transform is performed on each side  and using the quark-hadron duality principle, we equate both OPE and phenomenological sides. If a window of stability is obtained, the form factor and coupling constant can be extracted.  

In the following sects., we present the QCDSR formalism and our results for this calculation.

\section{Formalism}

The starting point to perform the sum rule, is the calculation of the three point correlation function of the vertex. In this case, the $J/\psi D^*_sD_s$ is a vector-vector-pseudoscalar ($VVP$) vertex. Each meson will be represented by an interpolating current, which contains the quantum numbers of the meson. 

We can calculate three correlation functions~\cite{Bracco20121019}: one with the vector meson $J/\psi$ off-shell, another with the vector meson $D_s^*$ off-shell and a third one with the pseudo-scalar meson $D_s$ off-shell:
\begin{align}
\Gamma^{(M)}(p,p') = \int d^4x d^4y  e^{ip'x}e^{-iqy} \Gamma^{(M)}(x,y) \label{eq:piMoff} 
\end{align}
where $q = p' - p$ is the transferred momentum, $M$ is the off-shell meson ($M=D_s,J/\psi,D^*_s$) and $\Gamma^{(M)}(x,y)$ is presented in eqs.~(\ref{eq:pijpsioff})-(\ref{eq:pidsoff}) below,  
\begin{eqnarray}
\Gamma^{(J/\psi)}_{\mu\nu}(x,y) &=& \bra{0'} T\{ j^{D_s}_5(x)j^{J/\psi\dagger}_\mu(y)j^{D_s^*\dagger}_\nu(0) \}\ket{0'}\,,  \label{eq:pijpsioff} \\
\Gamma^{(D^*_s)}_{\mu\nu}(x,y) &= & \bra{0'} T\{ j^{D_s}_5(x)j^{D^*_s\dagger}_\nu(y)j^{J/\psi\dagger}_\mu(0) \}\ket{0'}\,,\label{eq:pidsestoff}\\
\Gamma^{(D_s)}_{\mu\nu}(x,y) &= & \bra{0'} T\{ j^{D^*_s}_\nu(x)j^{D_s\dagger}_5(y)j^{J/\psi\dagger}_\mu(0) \}\ket{0'}\,,\label{eq:pidsoff}
\end{eqnarray}
where  $T$ is the time ordered product and $j_\mu^{J/\psi}$, $j_\nu^{D_s^*}$ and $j_5^{D_s}$ are the interpolating currents of the mesons $J/\psi$, $D_s^*$ and $D_s$ respectively.

According to the QCDSR, these correlation functions can be calculated in two different ways: using hadron degrees of freedom, called the \textit{phenomenological side}, or using quark degrees of freedom, called the \textit{OPE side}. In the following subsects., we calculate the correlations functions for the phenomenological and OPE sides, and the sum rule.

\subsection{The phenomenological side}
\label{sec:phen}

The effective Lagrangian, which represents the process in the $J/\psi D^*_s D_s$ vertex, is~\cite{Liu2007185,PhysRevC.69.035201}:
\begin{align}
\mathcal{L}_{J/\psi D^*_s D_s} = -g_{J/\psi D^*_s D_s} \varepsilon^{\alpha\beta\gamma\delta} \partial_\alpha \psi_\beta  \left ( \partial_\gamma D_{s\delta}^{*-} D^+_s + \partial_\gamma D_{s\delta}^{*+} D_s^- \right ),
\label{eq:lagrangeana}
\end{align}
where $\varepsilon^{0123} = +1$ is the Levi-Civita totally antisymmetric tensor.

From this Lagrangian, we can obtain the vertices of the hadronic process necessary to the calculation of the phenomenological side of the QCDSR. For the $J/\psi$, $D_s^*$ and $D_s$ off-shell cases they are, respectively:
\begin{align}
\braket{D_s^*(p)J/\psi(q)}{D_s(p')} &= 
i g_{J/\psi D^*_s D_s}^{(J/\psi)}(q^2)\epsilon_\beta(q,\lambda)\epsilon_\delta(p,\lambda)  q_\alpha p_\gamma \varepsilon^{\alpha\beta\gamma\delta}\,,\\
\braket{J/\psi(p)D_s^*(q)}{D_s(p')} &= 
i g_{J/\psi D^*_s D_s}^{(D^*_s)}(q^2)\epsilon_\beta(p,\lambda)\epsilon_\delta(q,\lambda) p_\alpha q_\gamma  \varepsilon^{\alpha\beta\gamma\delta}\,,\\
\braket{J/\psi(p)D_s(q)}{D^*_s(p')} &= 
-i g^{(D_s)}_{J/\psi D^*_s D_s}(q^2)\epsilon_\beta(p,\lambda)\epsilon^*_\delta(p',\lambda) p_\alpha p'_\gamma  \varepsilon^{\alpha\beta\gamma\delta}\,,
\end{align}
where $g^{(M)}_{J/\psi D^*_s D_s}(q^2)$ is the form factor of the $J/\psi D^*_s D_s$ vertex with meson $M$ off-shell ($M=J/\psi, D_s, D_s^*$). 

In the calculation of the phenomenological side, we also make use of the following hadronic matrix elements:
\begin{align} 
\bra{0} j_5^{P} \ket{P(q)} &= \bra{P(q)} j_5^{P} \ket{0} = f_{P} \frac{m^2_{P}}{m_{q_1} + m_{q_2}}\,,\\
\bra{V(q)} j_x^{V} \ket{0} &= f_{V} m_{V} \epsilon^*_x(q)\,, \\
\bra{0} j_x^{V} \ket{V(q)} &= f_{V} m_{V} \epsilon_x(q)\,,
\end{align}
where $P$ is the pseudo-scalar meson ($D_s$), $V$ is a vector meson ($J/\psi$ or $D_s^*$), $q$ is the four-momentum of the respective meson, $m^2_{P,V}$ is the squared mass of the meson, $f_{P,V}$ are the decay constants, $m_{q_1}$ and $m_{q_2}$ are the constituent quarks of meson $P$ and $\epsilon_{x}$ is the polarization vector, with $x = \mu(\nu)$ for $J/\psi (D_s^*)$.

Finally, the correlation functions (\ref{eq:pijpsioff})-(\ref{eq:pidsoff}) of the phenomenological side for each off-shell case read:
\begin{align}
\begin{split}
\Gamma^{phen (J/\psi)}_{\mu\nu} = \frac{C g^{(J/\psi)}_{J/\psi D^*_s D_s}(q^2) {p'}^\lambda p^\sigma \varepsilon_{\mu\nu\lambda\sigma} }{(p^2 + m_{D^*_s}^2)(q^2 + m_{J/\psi}^2)({p'}^2 + m_{D_s}^2)  } + h. r.\,,&\label{eq:fenomjpsioff}
\end{split}\\
\begin{split}
\Gamma^{phen (D^*_s)}_{\mu\nu} = \frac{-C g^{(D^*_s)}_{J/\psi D^*_s D_s}(q^2) {p'}^\lambda p^\sigma \varepsilon_{\mu\nu\lambda\sigma}}{(p^2 + m_{J/\psi}^2)(q^2 + m_{D^*_s}^2)({p'}^2 + m_{D_s}^2)} + h. r.\,,& \label{eq:fenomdsestoff}
\end{split}\\
\begin{split}
\Gamma^{phen (D_s)}_{\mu\nu} = \frac{C g^{(D_s)}_{J/\psi D^*_s D_s}(q^2) {p'}^\lambda p^\sigma \varepsilon_{\mu\nu\lambda\sigma} }{(p^2 + m_{J/\psi}^2)(q^2 + m_{D_s}^2)({p'}^2 + m_{D^*_s}^2)} + h. r.\,,& \label{eq:fenomdsoff}
\end{split}
\end{align}
where $h.r.$ stands for the contributions of higher resonances and continuum states of each meson and $C$ is defined as: 
\begin{align}
C = \frac{f_{D_s}f_{D^*_s} f_{J/\psi} m^2_{D_s}m_{D^*_s} m_{J/\psi}}{(m_c + m_s)}\,.
\label{eq:cteC}
\end{align}

From eqs.~(\ref{eq:fenomjpsioff})-(\ref{eq:fenomdsoff}), it is clear that this vertex has only one tensor structure to work within the formalism of the QCDSR.

\subsection{The OPE side}
\label{sec:ope}

The OPE side is obtained from eqs.~(\ref{eq:pijpsioff})-(\ref{eq:pidsoff}), with the interpolating currents written in terms of the quark fields: $j_\mu^{J/\psi} = \bar{c} \gamma_\mu c$, $j_\nu^{D_s^{*-}} = \bar{c} \gamma_\nu s$ and $j_5^{D_s^-} = i \bar{c} \gamma_5 s$. By construction, the OPE side is given by an expansion known as Wilson's Operator Product Expansion, dominated by the perturbative term and followed by non-perturbative contributions:
\begin{align}
\Gamma^{OPE(M)}_{\mu\nu} = \Gamma^{pert(M)}_{\mu\nu} + \Gamma^{non\mbox{-}pert(M)}_{\mu\nu} \,,
\label{eq:piladodaqcdgeral}
\end{align}
where $\Gamma^{pert(M)}_{\mu\nu}$ is the perturbative term and $\Gamma^{non\mbox{-}pert(M)}_{\mu\nu}$ are the non-perturbative contributions to the correlation function.  When calculating the form factor, this expansion exhibits a rapid convergence and can be truncated after a few terms \cite{Bracco20121019}. Considering the similarities between the $J/\psi D^*_s D_s$ and both $J/\psi D_s D_s$ and $J/\psi D^* D$ vertices, we expect, from the former, a similar behavior regarding the OPE series as seen in the two latter~\cite{OsorioRodrigues2014143,doi:10.1142/S0218301305003399}. Therefore, it should be adequate to consider non-perturbative contributions up to the mixed quark-gluon condensate: 
\begin{align}
\Gamma^{non\mbox{-}pert}_{\mu\nu}  = \Gamma^{\langle\bar{q}q\rangle}_{\mu\nu} + \Gamma^{m_q\langle\bar{q}q\rangle}_{\mu\nu} + \Gamma^{\langle g^2 G^2\rangle}_{\mu\nu} + \Gamma^{\langle \bar{q}g\sigma G q \rangle}_{\mu\nu} 
+ \Gamma^{m_q\langle \bar{q}g\sigma G q \rangle}_{\mu\nu} \,.
\label{eq:nonpertcontrib}
\end{align}

\begin{figure}[ht!]
\centering
 \subfigure[]{\includegraphics[width=0.27\linewidth]{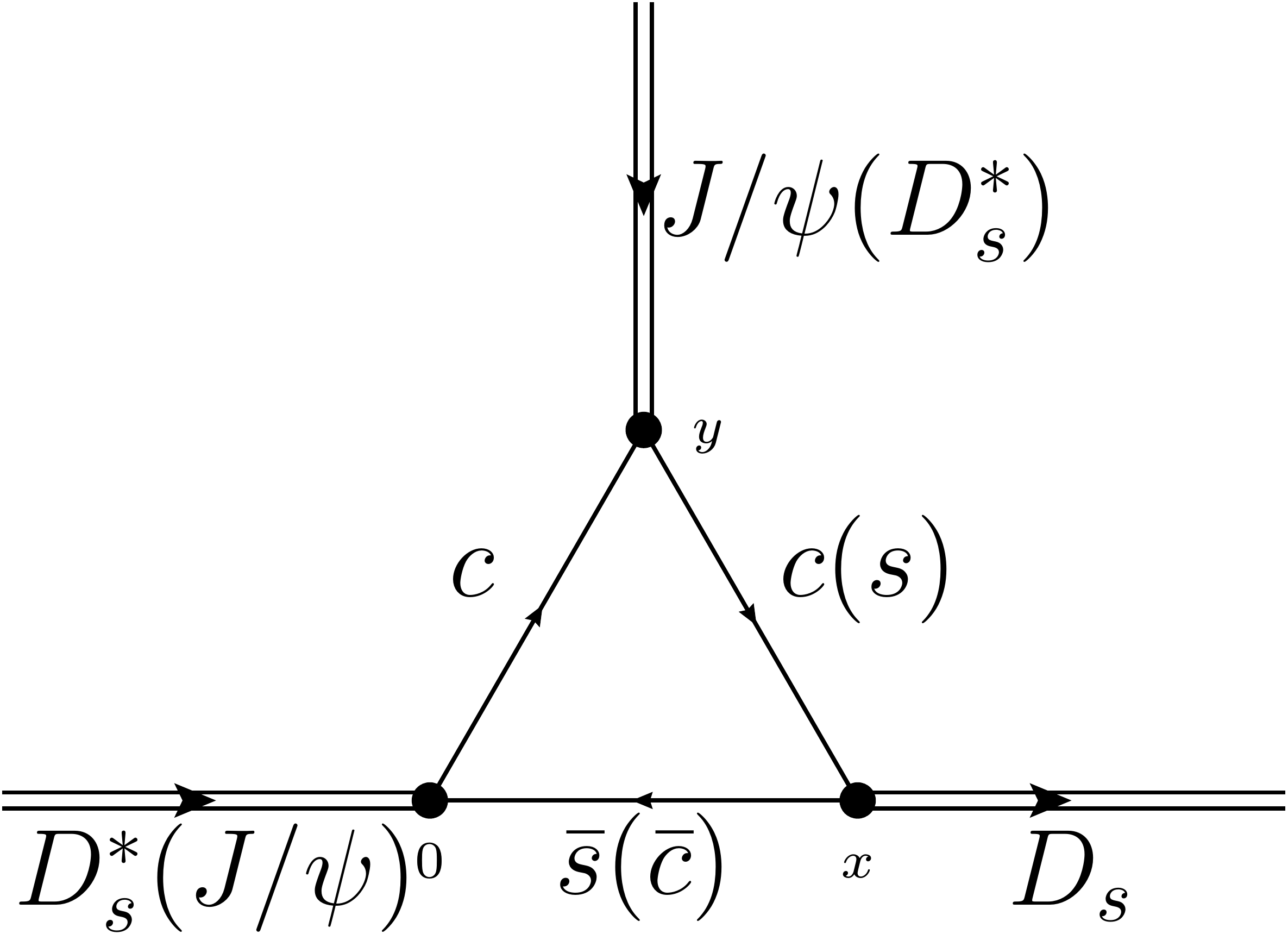} \label{subfig:a}}\quad
 \subfigure[]{\includegraphics[width=0.27\linewidth]{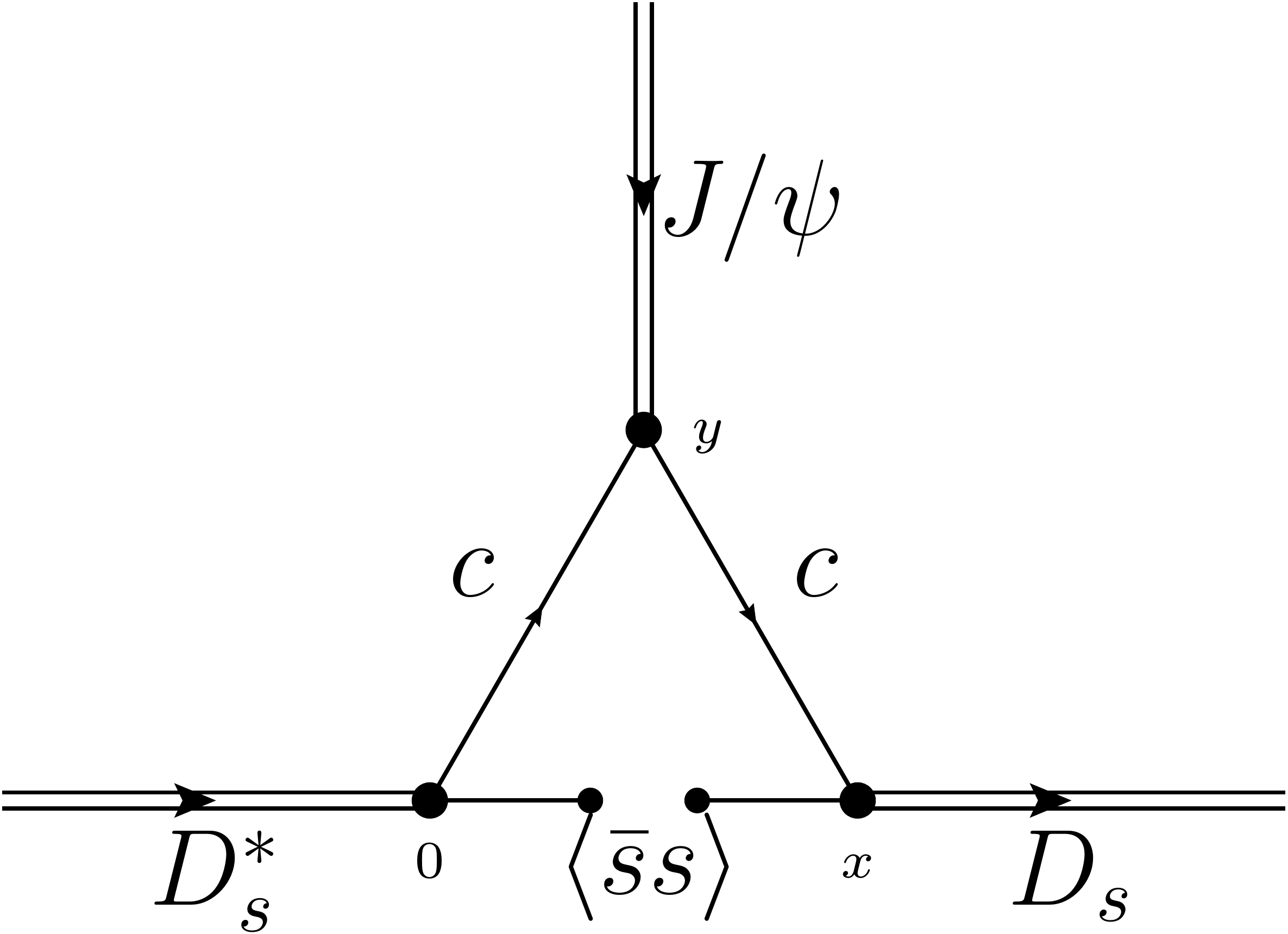}\label{subfig:b}}\quad
 \subfigure[]{\includegraphics[width=0.27\linewidth]{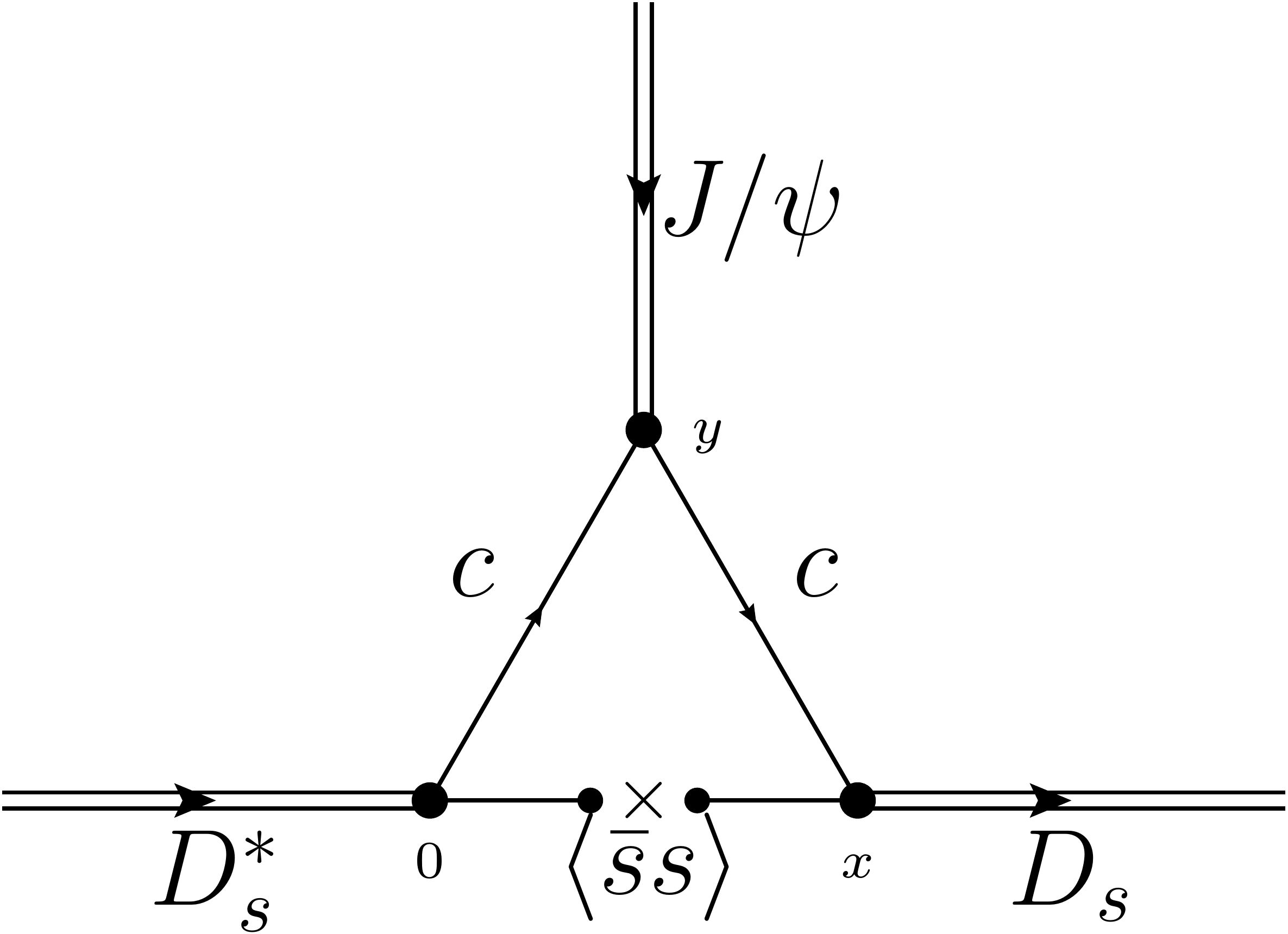}\label{subfig:c}}\\
 \subfigure[]{\includegraphics[width=0.27\linewidth]{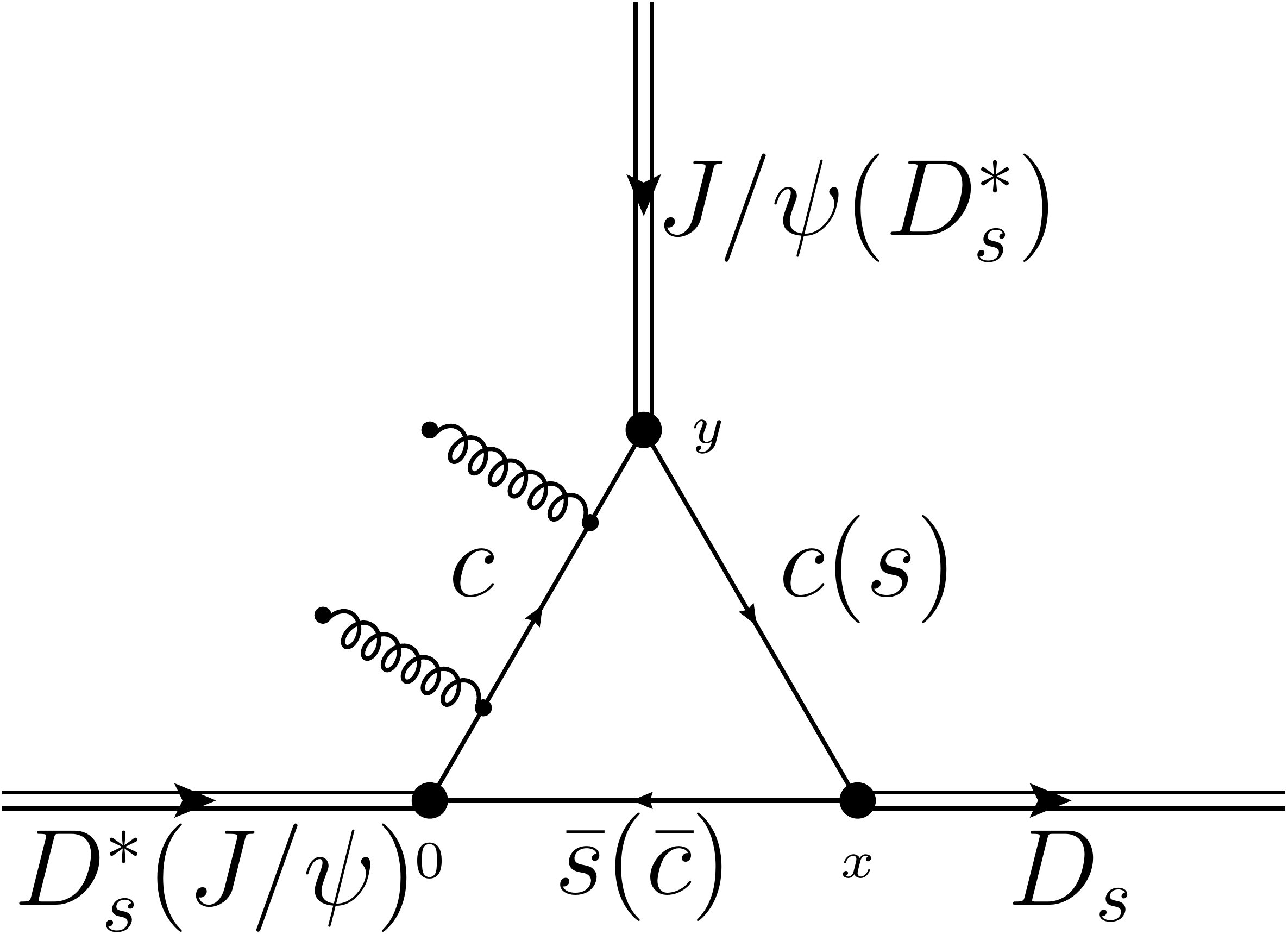}\label{subfig:d}}\quad
 \subfigure[]{\includegraphics[width=0.27\linewidth]{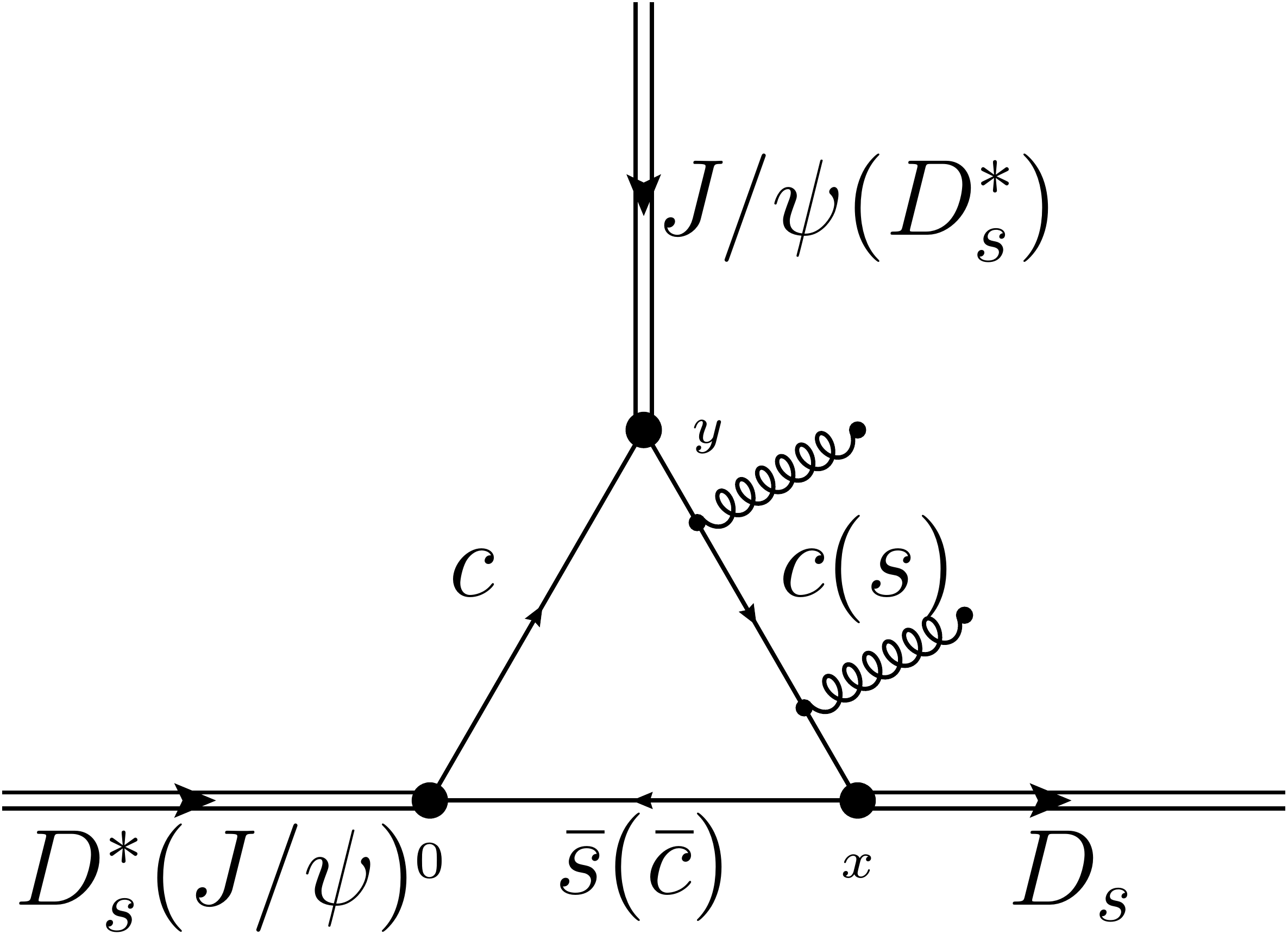}\label{subfig:e}}\quad
 \subfigure[]{\includegraphics[width=0.27\linewidth]{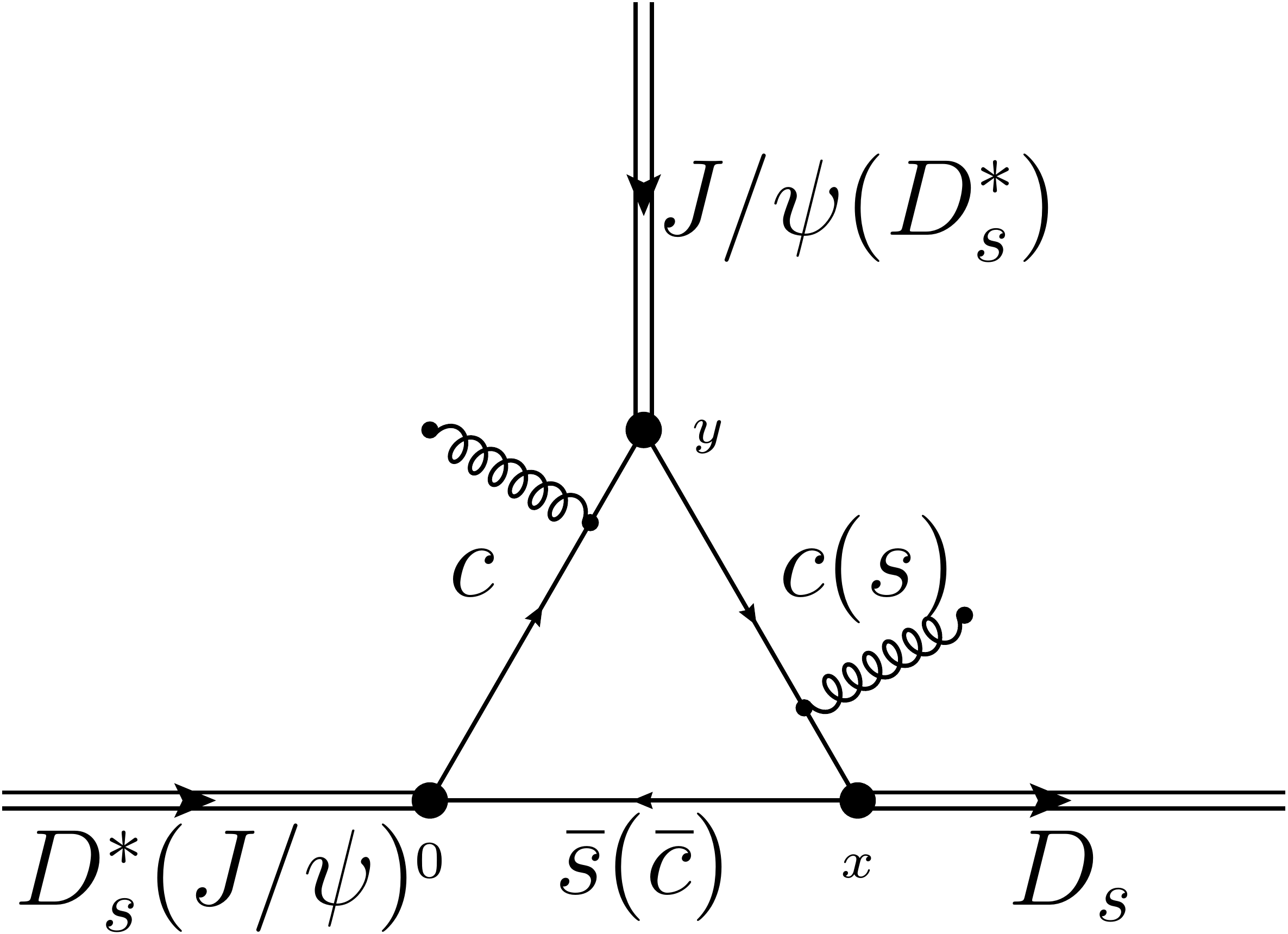}\label{subfig:f}} \\
 \subfigure[]{\includegraphics[width=0.27\linewidth]{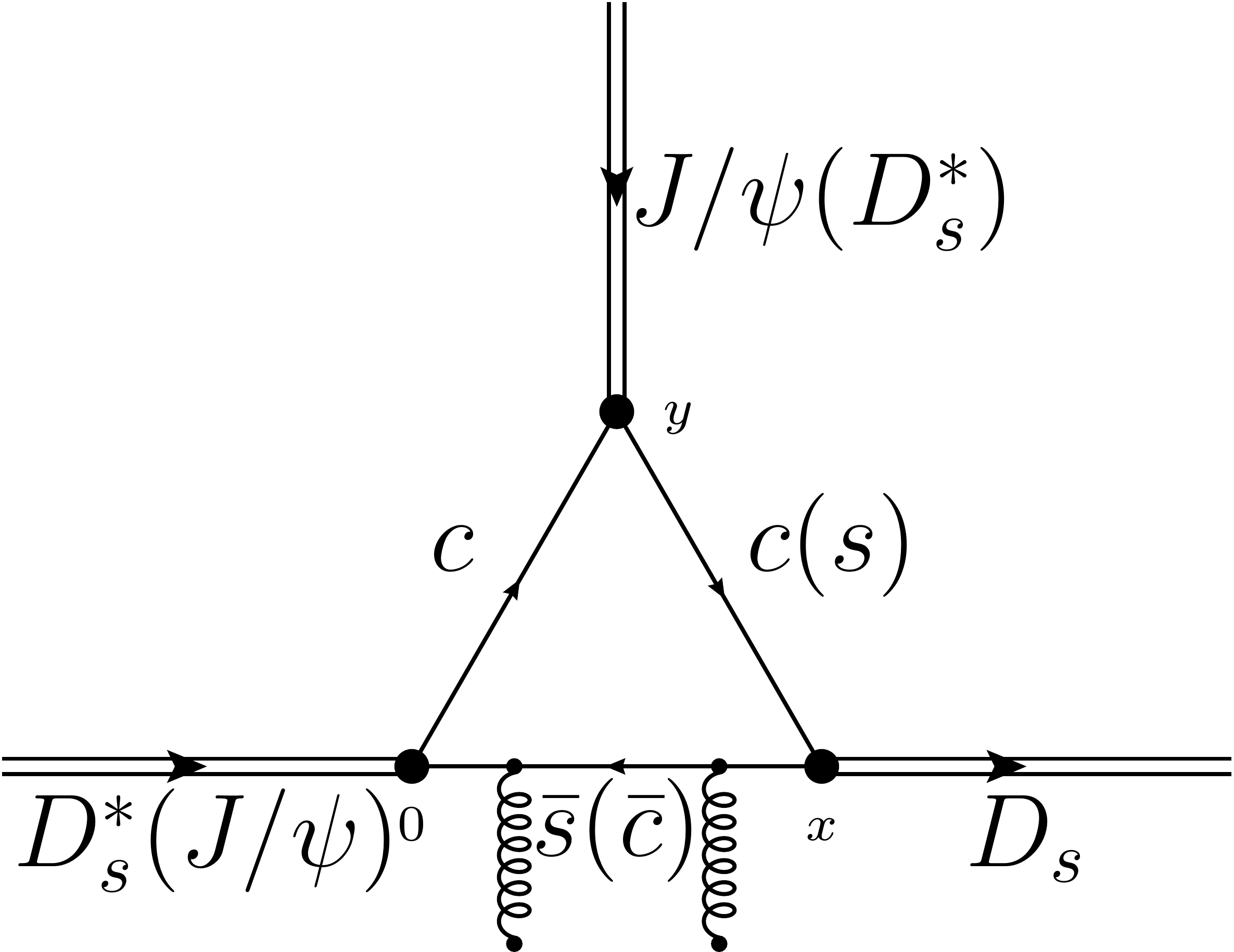}\label{subfig:g}}\quad
 \subfigure[]{\includegraphics[width=0.27\linewidth]{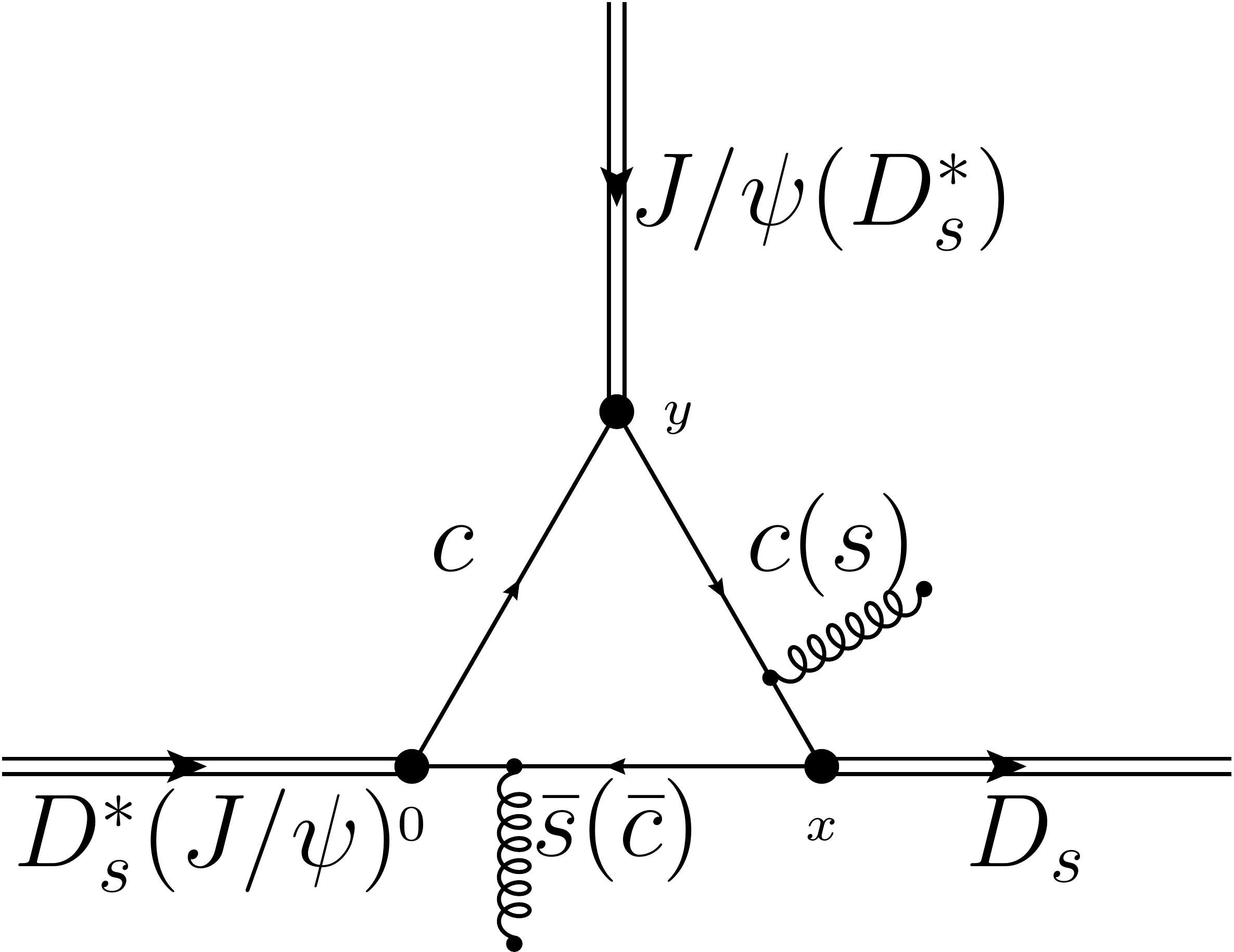}\label{subfig:h}}\quad
 \subfigure[]{\includegraphics[width=0.27\linewidth]{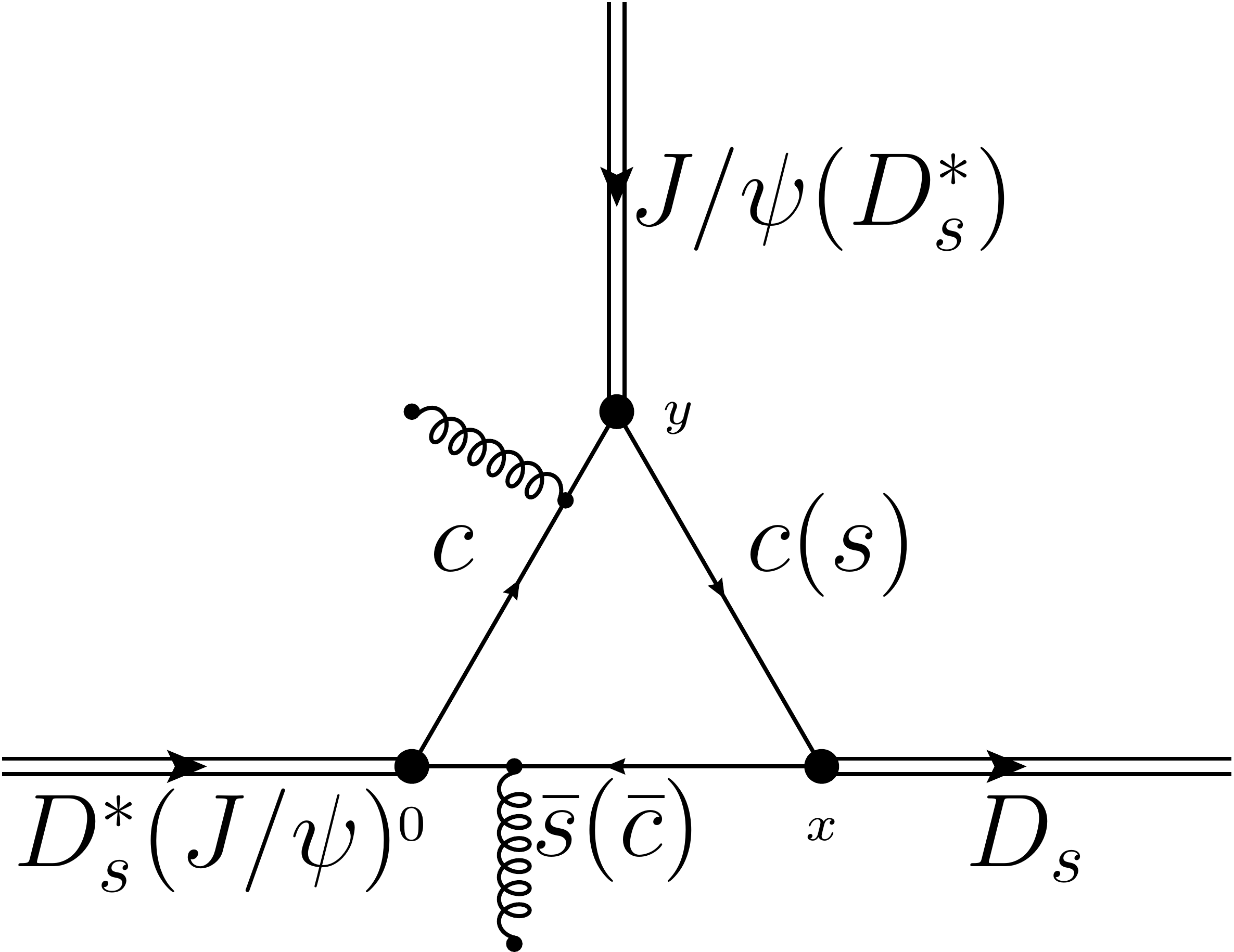}\label{subfig:i}} \\
 \subfigure[]{\includegraphics[width=0.27\linewidth]{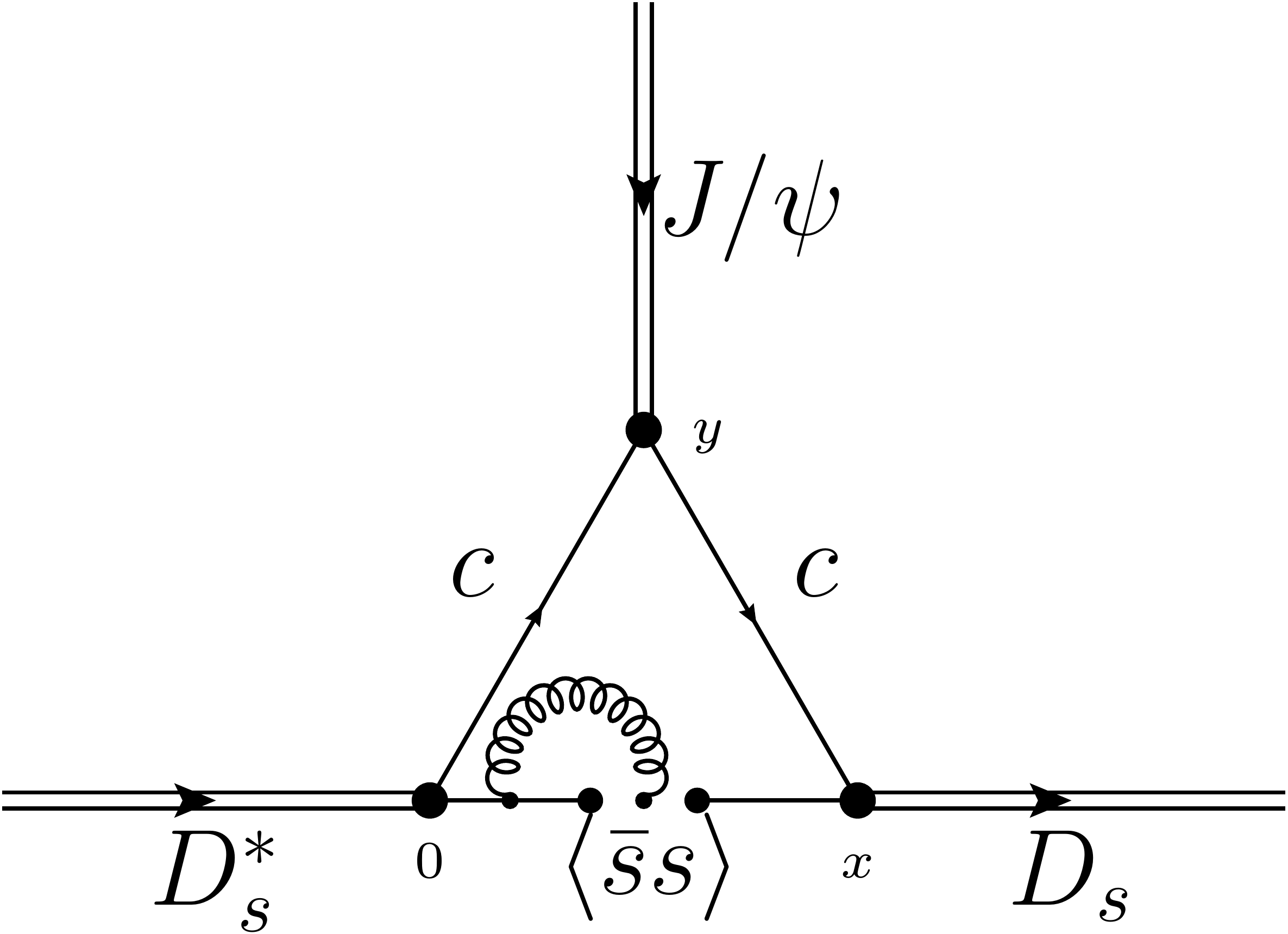}\label{subfig:j}}\quad
 \subfigure[]{\includegraphics[width=0.27\linewidth]{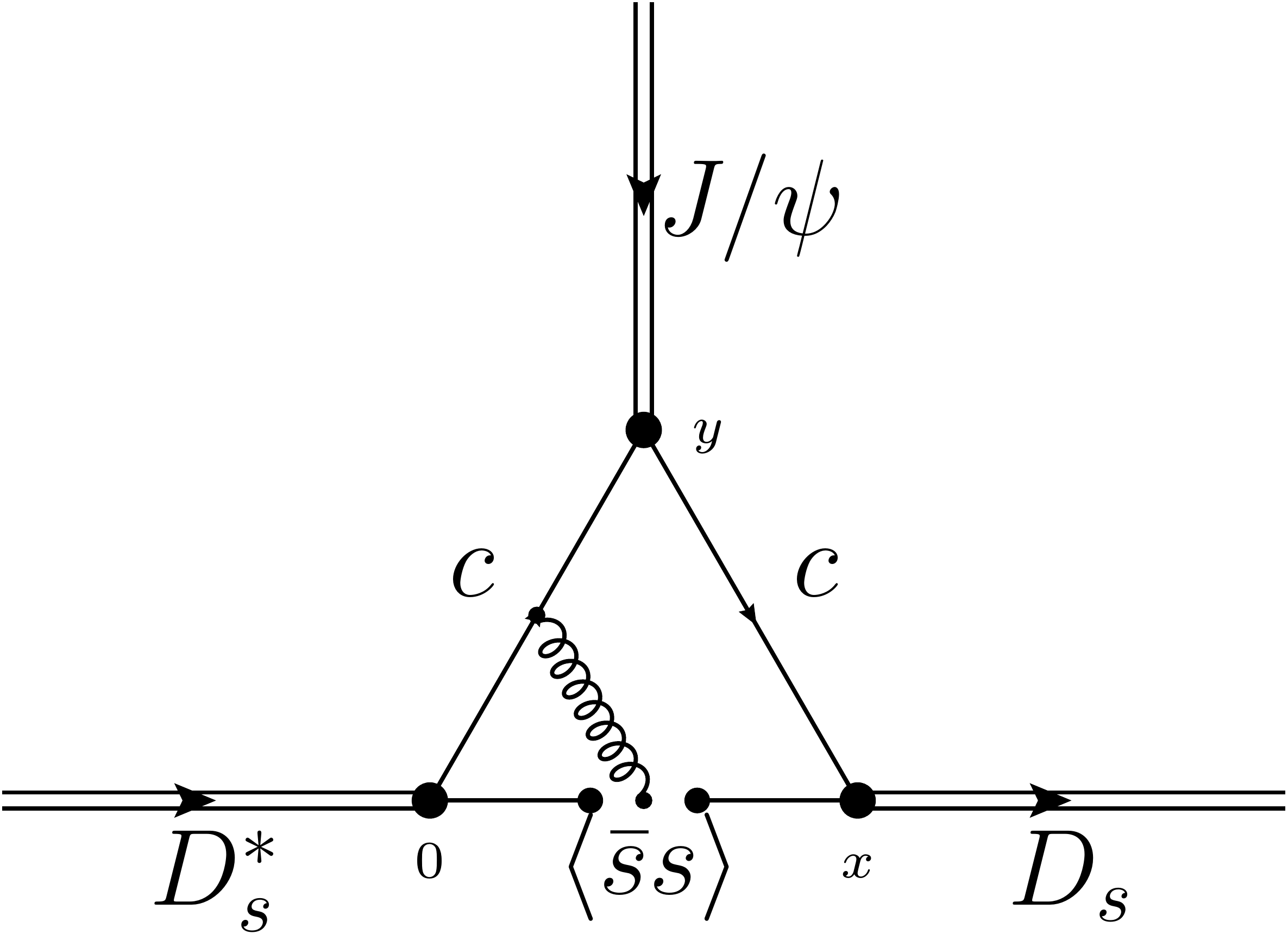}\label{subfig:k}} \quad
 \subfigure[]{\includegraphics[width=0.27\linewidth]{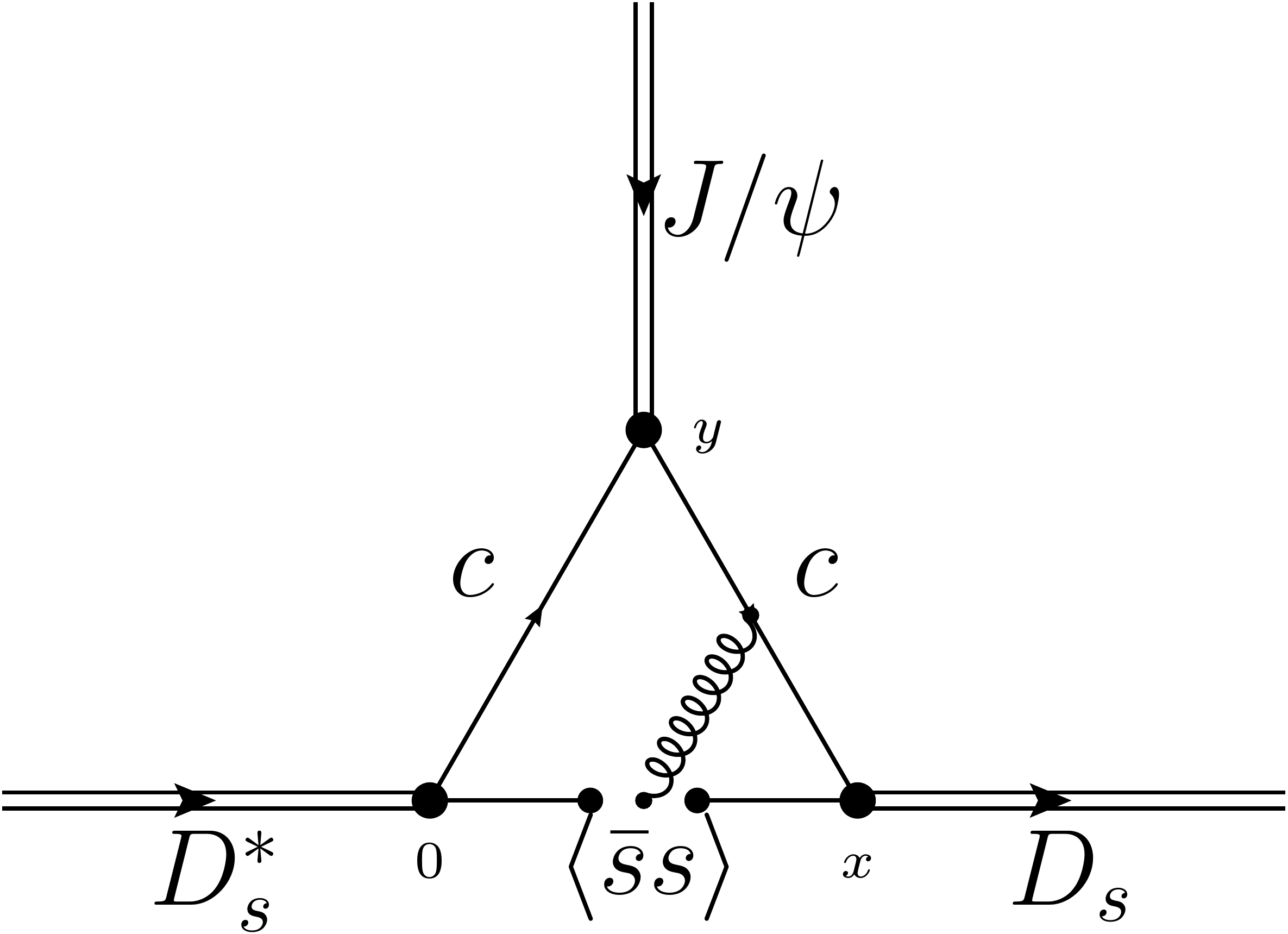}\label{subfig:l}} \\
 \subfigure[]{\includegraphics[width=0.27\linewidth]{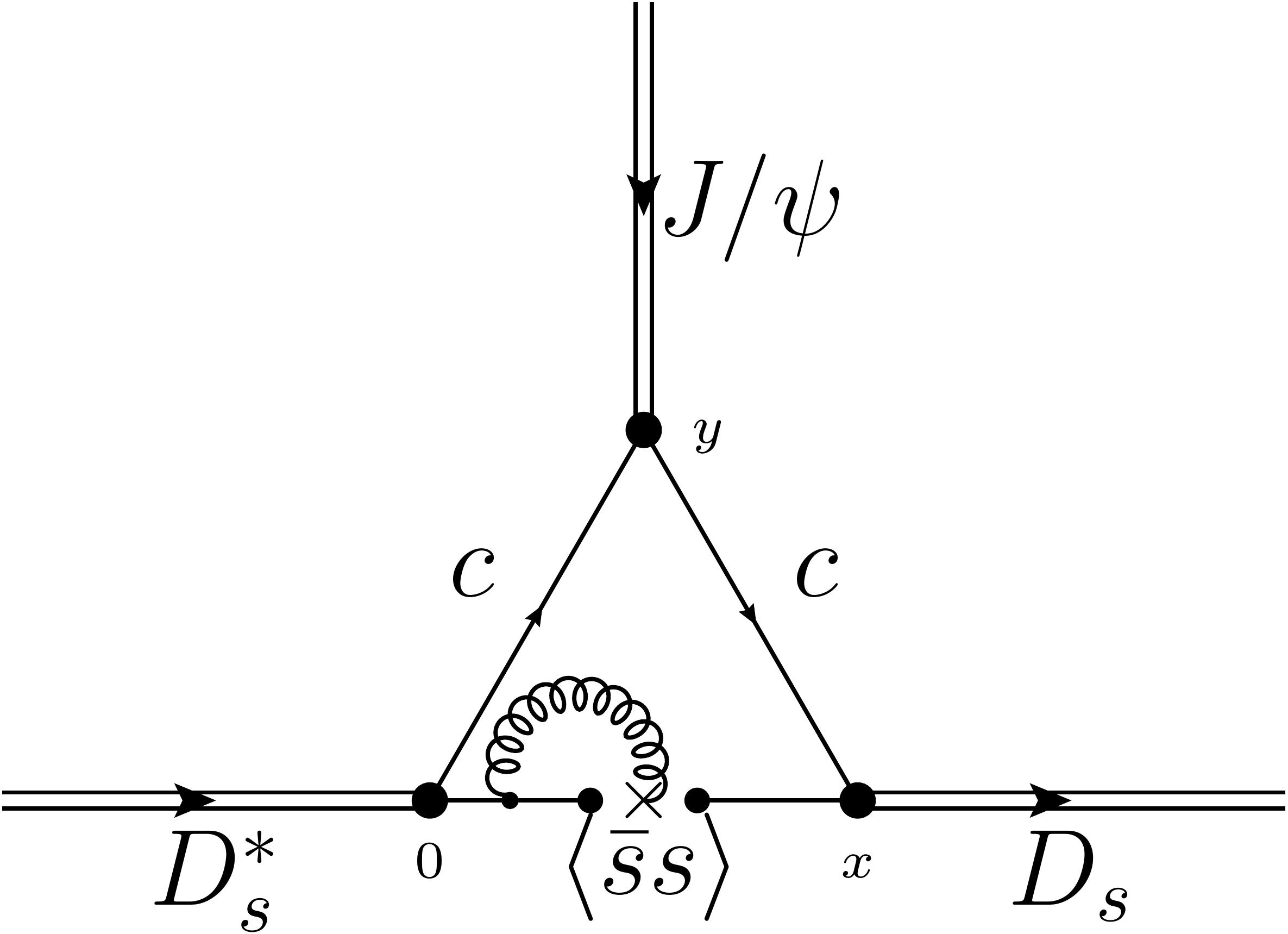}\label{subfig:m}} \quad
 \subfigure[]{\includegraphics[width=0.27\linewidth]{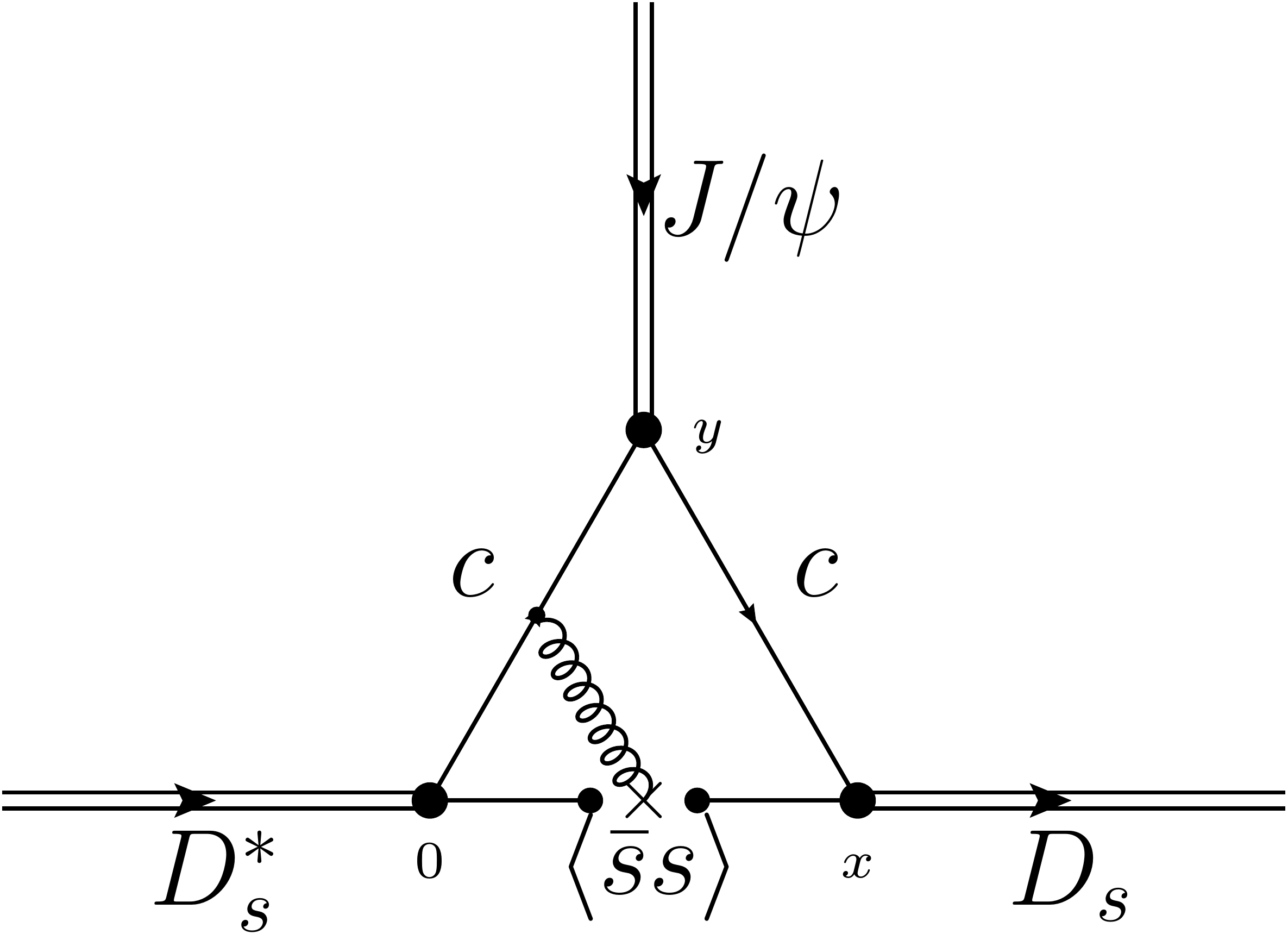}\label{subfig:n}}\quad
 \subfigure[]{\includegraphics[width=0.27\linewidth]{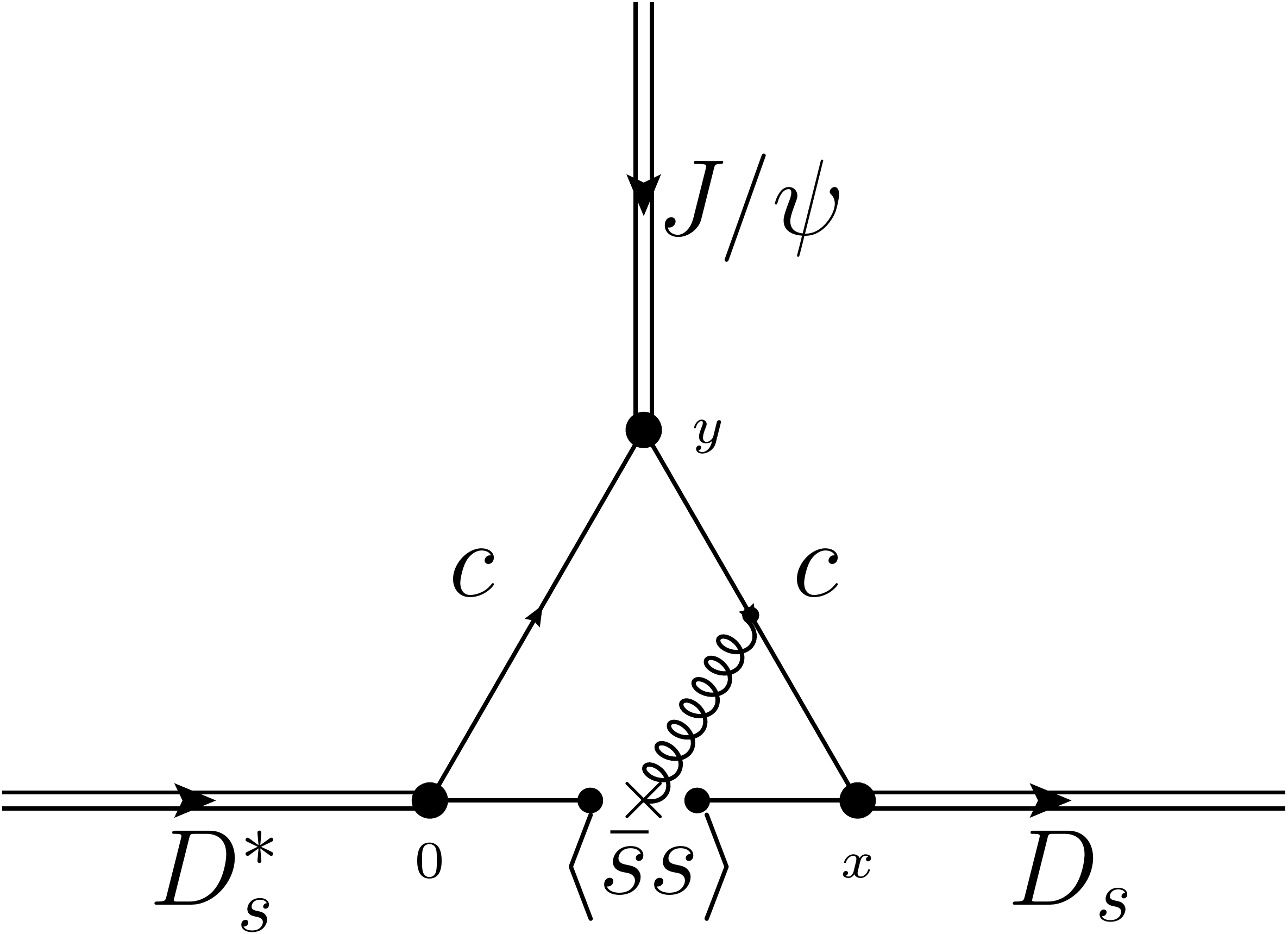}\label{subfig:o}}
 \caption{  \label{fig:diagrams}Contributing OPE diagrams for $J/\psi (D^{*}_s)$ off-shell. The $D_s$ off-shell case corresponds to a permutation of the $D_s$ and $D_s^*$ mesons of the $D_s^*$ off-shell case.}
\end{figure}

Calculating eq.~(\ref{eq:piladodaqcdgeral}) using the non-perturbative contributions of eq.~(\ref{eq:nonpertcontrib}) corresponds to calculating the diagrams of fig.~\ref{fig:diagrams}. As a consequence of the use of the double Borel transform, only the $J/\psi$ off-shell case has contributions from all the non-perturbative terms of eq.~(\ref{eq:nonpertcontrib}). All the non-perturbative terms, except the gluon condensates (fig.~\ref{subfig:d}-\ref{subfig:i}), are suppressed for both $D_s$ and $D_s^*$ off-shell cases. This has been taken into account in fig.~\ref{fig:diagrams}, wherefore these contributions were omitted from the beginning. The  $D_s$ off-shell case is also omitted in fig.~\ref{fig:diagrams}, as it is just an interchange between the mesons $D_s^*$ and $D_s$ of the $D_s^*$ off-shell case.

Using dispersion relations, the perturbative term (fig.~\ref{subfig:a}) for a given meson $M$ off-shell can be written in the following form:
\begin{align}
\Gamma^{pert (M)}_{\mu\nu}(p,p') = - \frac{1}{4\pi^2} \int^\infty_0 \int^\infty_0 \frac{\rho_{\mu\nu}^{pert (M)}(s, u, t)}{(s-p^2)(u-p'^2)} ds du \, ,
\label{eq:pertgeral}
\end{align}
where the spectral density $\rho_{\mu\nu}^{pert (M)}(s, u, t)$ can be obtained from the Cutkosky's rules. The quantities $s=p^2$, $u=p'^2$ and $t=q^2$ are the Mandelstam variables.

In general, eq.~(\ref{eq:pertgeral}) is the main contributing term of the OPE series in a QCDSR calculation. For a $VVP$ vertex, the spectral density  can be parametrized as:
\begin{align}
\rho_{\mu\nu}^{pert (M)}(s,u,t) = \frac{3}{\sqrt{\lambda}} F^{(M)}(s,u,t) {p'}^\lambda p^\sigma \varepsilon_{\mu\nu\lambda\sigma} \, ,
\label{dspec}
\end{align}
where $\lambda = (u + s - t)^2 - 4us$ and $F^{(M)}$ is an invariant amplitude. For the $ J/\psi$  and $D_s$ off shell, the invariant amplitude can be written as:
\begin{eqnarray}
F^{(J/\psi)} &=& (m_s-m_c)(A+B) - m_s \, ,\\
F^{(D_s)} &=& -F^{(D^*_s)} = (m_c - m_s)B - m_c \, ,
\end{eqnarray}
where 
\begin{align}
&A = \left [ \frac{\bar{k}_0}{\sqrt{s}} - \frac{p'_0 \overline{|\vec{k}|} \overline{\cos\theta}}{|\vec{p'}|\sqrt{s}} \right ] \, ,
\;\;\;\;\;\;\;\;\;\;\;\;
B = \frac{\overline{|\vec{k}|} \overline{\cos\theta}}{|\vec{p'}|} \, ,
\nonumber \\
&\;\;\;\;\;\;\;\;\;\;\;\;\overline{\cos\theta} = \frac{2p'_0\bar{k}_0 - u + (-1)^{(\epsilon)}(m_s^2 - m_c^2) }{2|\vec{p'}|\overline{|\vec{k}|}} \, ,
\nonumber \\
&\overline{|\vec{k}|} = \sqrt{\bar{k}_0^2 - \epsilon m_s^2 - (1-\epsilon)m_c^2} \, ,
\;\;\;\;\;\;\;\;
\bar{k}_0 = \frac{s + \epsilon(m_s^2-m_c^2)}{2\sqrt{s}} \, ,
\nonumber \\
&\;\;\;\;\;\;\;\;\;\;\;\;p'_0 = \frac{s+u-t}{2\sqrt{s}} \, ,\;\;\;\;\;\;\;\;|\vec{p'}| = \frac{\sqrt{\lambda}}{2\sqrt{s}}\, , \nonumber
\end{align}
and $\epsilon = 0(1)$ for $D^{(*)}_s(J/\psi)$ off-shell.

The first non-perturbative contribution to the correlation function (\ref{eq:piladodaqcdgeral}) is the quark condensate showed in 
fig.~\ref{subfig:b}: 
\begin{align}
\Gamma^{\langle\bar{s}s\rangle (J/\psi)}_{\mu\nu} = \frac{-\langle\bar{s}s\rangle {p'}^\lambda p^\sigma \varepsilon_{\mu\nu\lambda\sigma} }{(p^2 - m^2_c)(p'^2 - m_c^2)}\,.
\label{eq:contribcond}
\end{align}

Figure~\ref{subfig:c} is the first order mass term for the strange quark condensate contribution to eq.~(\ref{eq:piladodaqcdgeral}), numerically less important and usually negligible in the three point QCDSR:
\begin{align}
\Gamma^{m_s\langle\bar{s}s\rangle (J/\psi)}_{\mu\nu} = \frac{m_c m_s\langle\bar{s}s\rangle \left ( {p'}^2+p^2-2m_c^2 \right )}{2\left (p^2 - m_c^2 \right )^2 \left ({p'}^2 - m_c^2 \right )^2}{p'}^\lambda p^\sigma \varepsilon_{\mu\nu\lambda\sigma}\,.
\end{align}

Full expressions for the gluon condensates ($\langle g^2 G^2 \rangle$) of fig.~\ref{subfig:d}-\ref{subfig:i} and mixed quark-gluon condensates ($\langle \bar{s}g\sigma \cdot G s \rangle$) of 
fig.~\ref{subfig:j}-\ref{subfig:o} for the $J/\psi$ off-shell case can be found in Appendix~\ref{appendix:remaningexpressions}. 

\subsection{The QCD Sum Rule}
\label{sec:rsqcd}

Two more steps are necessary in order to calculate the QCDSR to obtain the form factors. 
First, we make the change of variables $p^2 \rightarrow -P^2$,  $p'^2 \rightarrow -{P'}^2$ and $q^2 \rightarrow -Q^2$ followed by a double Borel transform~\cite{10.1142/9789812776310_0004,doi:10.1142/9789812810458_0033} to both sides of the QCDSR in eqs.~(\ref{eq:fenomjpsioff})-(\ref{eq:fenomdsoff}) and~(\ref{eq:piladodaqcdgeral}), which involves the transformation: $P^2  \to M^2$ and $P'^2  \to M'^2$, where $M$ and $M'$ are the Borel masses. After that, we equate the phenomenological and OPE sides,  invoking the quark-hadron duality from which the sum rule is obtained.

The second step is to eliminate the $h.r.$ terms from the phenomenological side in eqs.~(\ref{eq:fenomjpsioff})-(\ref{eq:fenomdsoff}). This is achieved by the introduction of the continuum threshold parameters $s_0$ and $u_0$ in the OPE side and by taking advantage of the quark-hadron duality. These parameters fulfill the following relations: $m_i^2 < s_0 < {m'}_i^2$ and $m_o^2 < u_0 < {m'}_o^2$,  where  $m_i$ and $m_o$ are the masses of the incoming and outcoming mesons respectively and $m'$ is the mass of the first excited state of these mesons.

After performing these steps, we obtain the QCDSR expressions for the form factors:
\begin{widetext}
\begin{align}
\label{eq:rsqcdjpsi}
g_{J/\psi D^*_s D_s}^{(J/\psi )}(Q^2) & = \frac{-\frac{3}{4\pi^2} \int^{s_0}_{s_{inf}} \int^{u_0}_{u_{inf}} \frac{1}{\sqrt{\lambda}}F^{(J/\psi)} e^{-\frac{s}{M^2}}e^{-\frac{u}{M'^2}} ds du + \mathcal{B}\mathcal{B}\left [ \Gamma^{non\mbox{-}pert} \right]}{\frac{C}{(Q^2+m_{J/\psi }^2)}e^{-m_{D^*_s}^2/M^2}e^{-m_{D_s}^2/M'^2}}\;,\\
\label{eq:rsqcddsest}
g_{J/\psi  D^*_s D_s}^{(D^*_s)}(Q^2) & = \frac{\frac{3}{4\pi^2} \int^{s_0}_{s_{inf}} \int^{u_0}_{u_{inf}} \frac{1}{\sqrt{\lambda}}F^{(D^*_s)} e^{-\frac{s}{M^2}}e^{-\frac{u}{M'^2}} ds du + \mathcal{B}\mathcal{B}\left [ \Gamma^{\langle g^2 G^2\rangle}  \right]}{\frac{C}{(Q^2+m_{D^*_s}^2)} e^{-m_{J/\psi }^2/M^2}e^{-m_{D_s}^2/M'^2}}\;,\\
\label{eq:rsqcdds}
g_{J/\psi  D^*_s D_s}^{(D_s)}(Q^2) & = \frac{-\frac{3}{4\pi^2} \int^{s_0}_{s_{inf}} \int^{u_0}_{u_{inf}} \frac{1}{\sqrt{\lambda}}F^{(D_s)} e^{-\frac{s}{M^2}}e^{-\frac{u}{M'^2}} ds du + \mathcal{B}\mathcal{B}\left [ \Gamma^{\langle g^2 G^2\rangle}  \right]}{\frac{C}{(Q^2+m_{D_s}^2)} e^{-m_{J/\psi }^2/M^2}e^{-m_{D^*_s}^2/M'^2}}\;,
\end{align}
\end{widetext}
where $\mathcal{B}\mathcal{B}\left [ \;\;\; \right ]$ stands for the double Borel transform and $\mathcal{B}\mathcal{B}\left [ \Gamma^{non\mbox{-}pert} \right] = \mathcal{B}\mathcal{B}\left [ \Gamma^{\langle g^2 G^2\rangle} \right]$ for the $D^{(*)}_s$ off-shell cases, as already mentioned. The constant $C$ is defined in eq.~(\ref{eq:cteC}).

As in previous works, the definition for the coupling constant $g_{J/\psi D^*_s D_s}$ is given by~\cite{Bracco20011}:
\begin{align} 
\lim\limits_{Q^2 \to -m_M^2} g^{(M)}_{J/\psi D^*_s D_s}(Q^2)\;.
\label{eq:defctedeacoplamento}
\end{align}
In order to calculate this, it is necessary to extrapolate the results for the form factor to the region of $Q^2<0$. From eqs.~(\ref{eq:rsqcdjpsi})-(\ref{eq:rsqcdds}) it is clear that we can obtain the coupling constant from three different form factors, one for each meson off-shell. However, the coupling constant must be the same regardless the form factor used for the extrapolation. This condition is used to minimize the uncertainties in  the calculation of the coupling constant, which will become quite clear in the following section.

\section{Results and Discussion}
Equations~(\ref{eq:rsqcdjpsi})-(\ref{eq:rsqcdds}) show the three different form factors of the $J/\psi D_s^* D_s$ vertex. The numerical calculation of these form factors give results that must be fitted to an analytical function of $Q^2$.  In order to minimize the uncertainties regarding the fitting procedure, it is required that these three form factors lead to the same coupling constant~\cite{Bracco20011}. This condition also helps to reduce the errors of finding the Borel masses and continuum thresholds necessary to the calculation.

Regarding the Borel masses, which satisfy the relation $M'^2 = \frac{m_o^2}{m_i^2}M^2$, they can assume any value within the Borel window of stability, in which the phenomenologial and OPE side should be compatible. The Borel window can be reduced by requiring  that the pole contribution must be bigger than the continuum contribution by at least $50 \%$ and that the perturbative term contributes with more than $50\%$ to the total correlation function.  We have worked with the average values of the form factors within the Borel window. Hence, it is not necessary to use any specific Borel mass value, minimizing the uncertainties associated to the choice of this quantity \cite{OsorioRodrigues2014143}. The procedure is briefly sketched as follows: the average value of the form factor is calculated in the Borel window for each value of $Q^2$ used.  The standard deviation is then used to automatize the analysis of the stability of the form factor regarding the Borel masses and continuum threshold parameters. This stability criterion then indicates the optimal values of the continuum threshold parameters and the Borel windows we have to work with. Thus, not only is a good stability ensured in the Borel window but also in the whole $Q^2$ interval, without the need to chose a specific value for the Borel mass.

\begin{figure}[!ht]
\centering
 \subfigure[]{\includegraphics[width=0.49\linewidth]{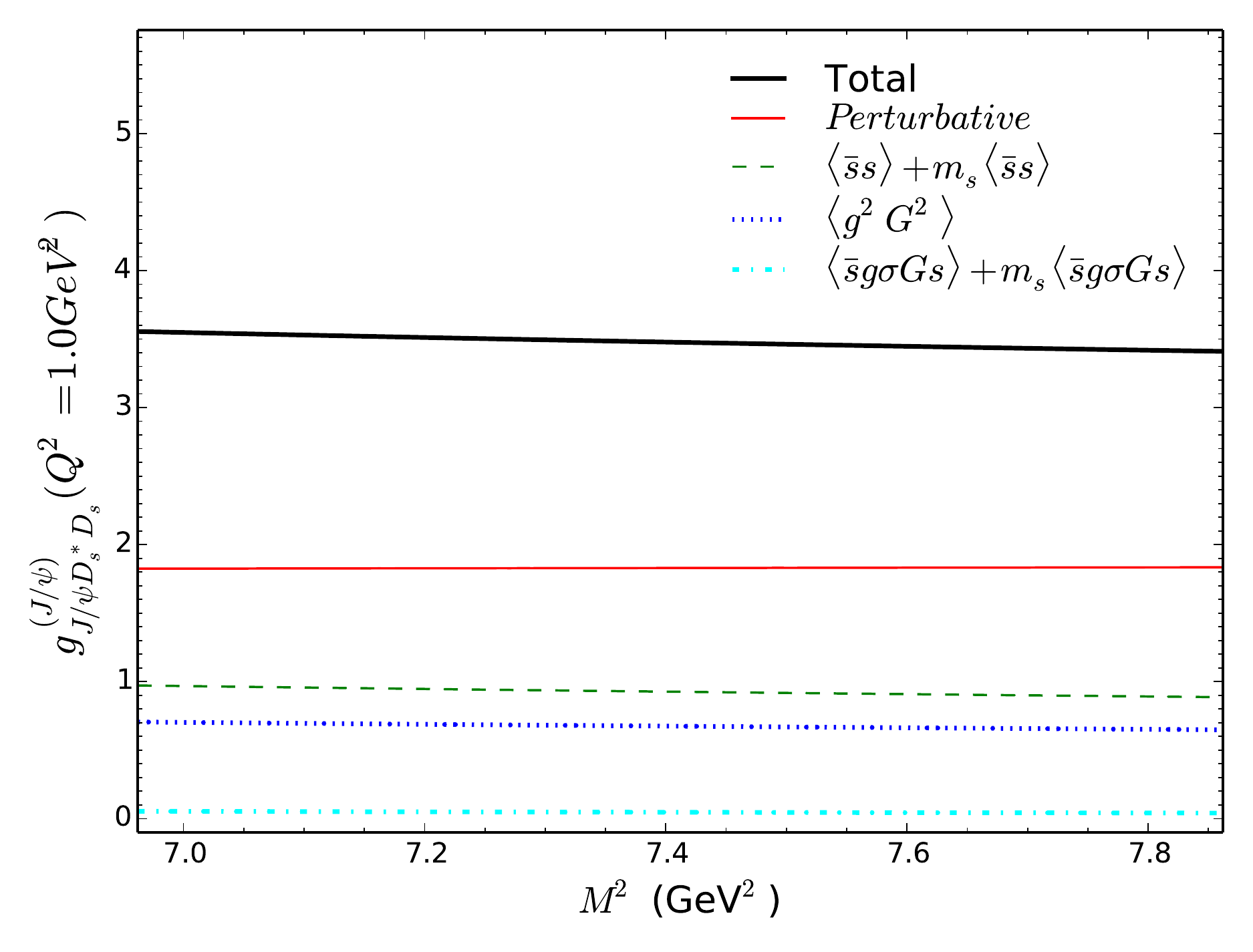} \label{subfig:contrib_jpsi}}
 \subfigure[]{\includegraphics[width=0.49\linewidth]{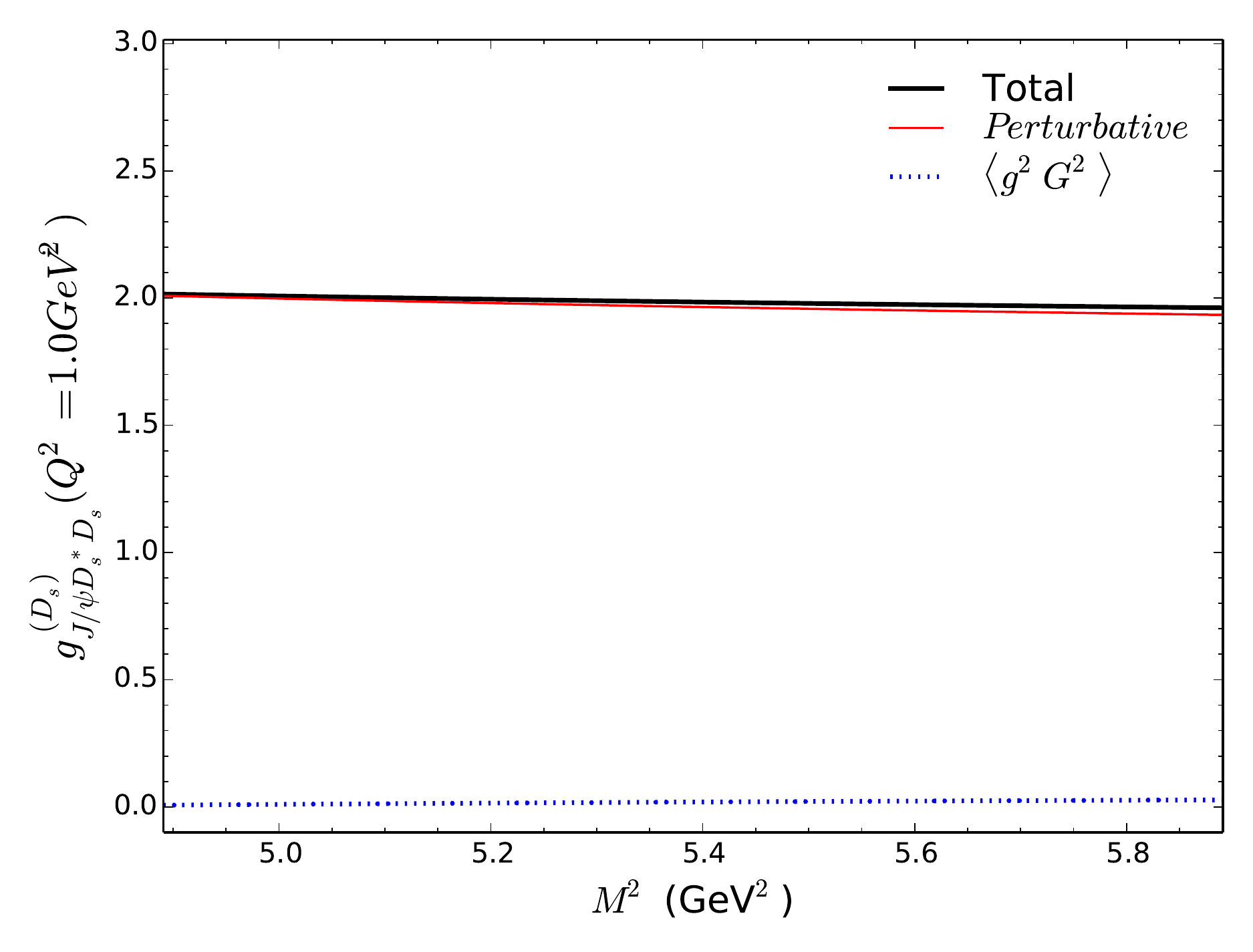}\label{subfig:contrib_ds}}
  \caption{\label{fig:estabilidade}OPE contributions for $J/\psi$ off-shell (panel \textbf{(a)}) and $D_s$ off-shell (panel \textbf{(b)}).}
  \end{figure}

The continuum threshold parameters, $s_0$ and $u_0$, are defined as $s_0 = (m_i + \Delta_i)^2$ and $u_0 = (m_o + \Delta_o)^2$, where  the quantities $\Delta_i$ and $\Delta_o$ have been determined imposing the most stable Borel window, as mentioned before.  In order to include the pole and to exclude the contributions from higher resonances and continuum states, the values for $ \Delta_{J/\psi}$, $\Delta_{D^*_s}$ and $\Delta_{D_s}$ cannot be far from the experimental value (when available) of the distance between the pole and the first excited state~\cite{PDG,PhysRevD.84.034006}. Our analysis has found that the best values are $\Delta_{D_s} = 0.6$ GeV and $\Delta_{D^*_s} = \Delta_{J/\psi} = 0.5$ GeV, which leads to remarkably stable Borel windows for the three off-shell cases, as it can be seen in fig.~\ref{fig:estabilidade} for $J/\psi$ and $D_s$ off-shell. Figures for $D_s^*$ off-shell are omitted as they are very similar to the $D_s$ off-shell case.

Also in fig.~\ref{fig:estabilidade}, we can check that the perturbative term is the leading one in the OPE series, followed by the quark condensate (for the $J/\psi$ off-shell case) and the gluon condensates. The summed contribution of the non-perturbative terms, $m_s \langle \bar{s}s \rangle$, $\langle \bar{s}g\sigma G s \rangle$ and $m_s \langle \bar{s}g\sigma G s \rangle$, is small ($\approx 2\%$), which means we could easily neglect these terms without changing our results significantly. Likewise, the gluon condensates have a very small contribution to the total correlation function in the $D_s$ and $D_s^*$ off-shell cases and they may also be neglected.

Besides the values of the Borel masses and the continuum  thresholds, we also need to know the values of decay constants,  condensates and hadrons and quark masses. The hadron masses used are $m_{D_s} = 1.968$ GeV, $m_{D^*_s} = 2.112$ GeV and $m_{J/\psi} = 3.097$ GeV ~\cite{PDG}. The other parameters are presented in table~\ref{tab:errors}.


In table~\ref{tab:results}, we present the $Q^2$ and Borel windows found for each form factor $g^{(M)}_{J/\psi D^*_sD_s}(Q^2)$ (where $M$ is the meson off-shell), together with its parametrization and the corresponding coupling constant $g_{J/\psi D^*_sD_s}$. Its associated error $\sigma$ is calculated using the method that will be explained later in this work.
\begin{table}[ht]
\caption{\label{tab:results}Parametrization of the form factors and numerical results for the coupling constant of this work.}
\begin{tabular}{cccc} 
\hline
\hline
Quantity                      & $J/\psi$ off-shell & $D^*_s$ off-shell & $D_s$ off-shell \\ 
\hline
$Q^2$ (GeV$^2$)            & [0.5, 2.0]  & [1.0, 3.0]  & [1.0, 4.0]   \\
$M^2$ (GeV$^2$)            & [7.0, 7.9]  & [5.3, 7.3]  & [4.9, 5.9]   \\
$g^{(M)}_{J/\psi D^*_sD_s}(Q^2)$ & $\frac{A}{B+Q^2}$ & $A e^{-Q^2/B}$ & $A e^{-Q^2/B}$\\
$A$                          & 193.4 GeV & 2.003 GeV$^{-1}$ & 2.330 GeV$^{-1}$       \\
$B$ (GeV$^{2}$)             & 54.46     & 6.027  & 6.122       \\
${g^{(M)}_{J/\psi D^*_sD_s}}\pm\sigma$ (GeV$^{-1}$)     & $4.31^{+0.41}_{-0.38}$ & $4.20^{+0.22}_{-0.18}$          & $4.39^{+0.27}_{-0.23}$  \\
\hline
\hline
\end{tabular}
\end{table}

\begin{figure}[ht]
 \subfigure[]{\includegraphics[width=0.48\linewidth]{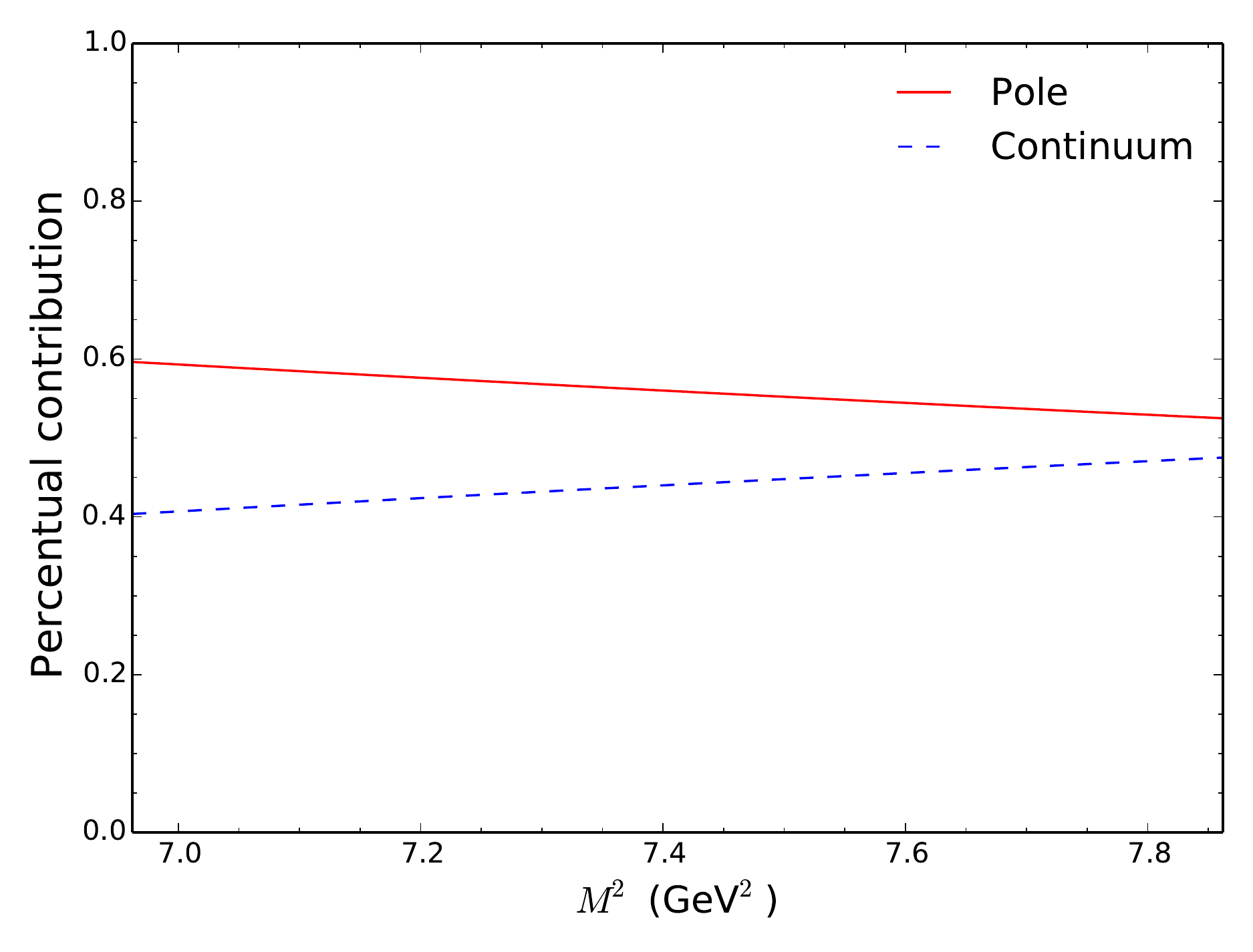} \label{subfig:polo_jpsi}}
 \subfigure[]{\includegraphics[width=0.48\linewidth]{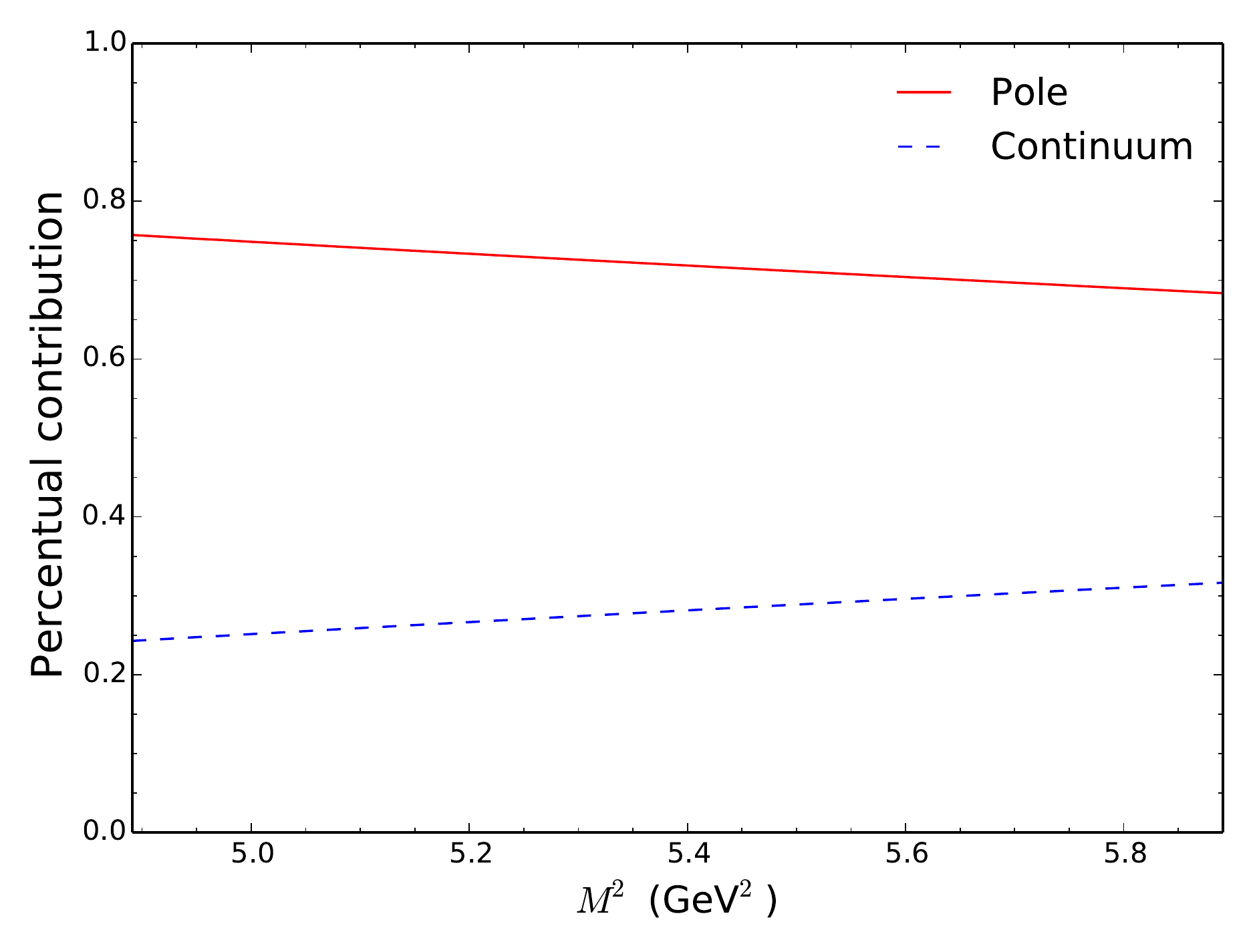}\label{subfig:polo_ds}}
  \caption{\label{fig:jpsidsestdspolocont}Pole and continuum contributions for the $J/\psi$ off-shell (panel \textbf{(a)}) and for $D_s$ off-shell (panel \textbf{(b)}), both at $Q^2 = 1$ GeV$^2$.}
  \end{figure}

The Borel windows presented in table~\ref{tab:results} ($M^2$ row) satisfy the already mentioned conditions regarding the pole and continuum contributions, as shown in fig.~\ref{fig:jpsidsestdspolocont}. 

\begin{figure}[ht]
  \includegraphics[width=\linewidth]{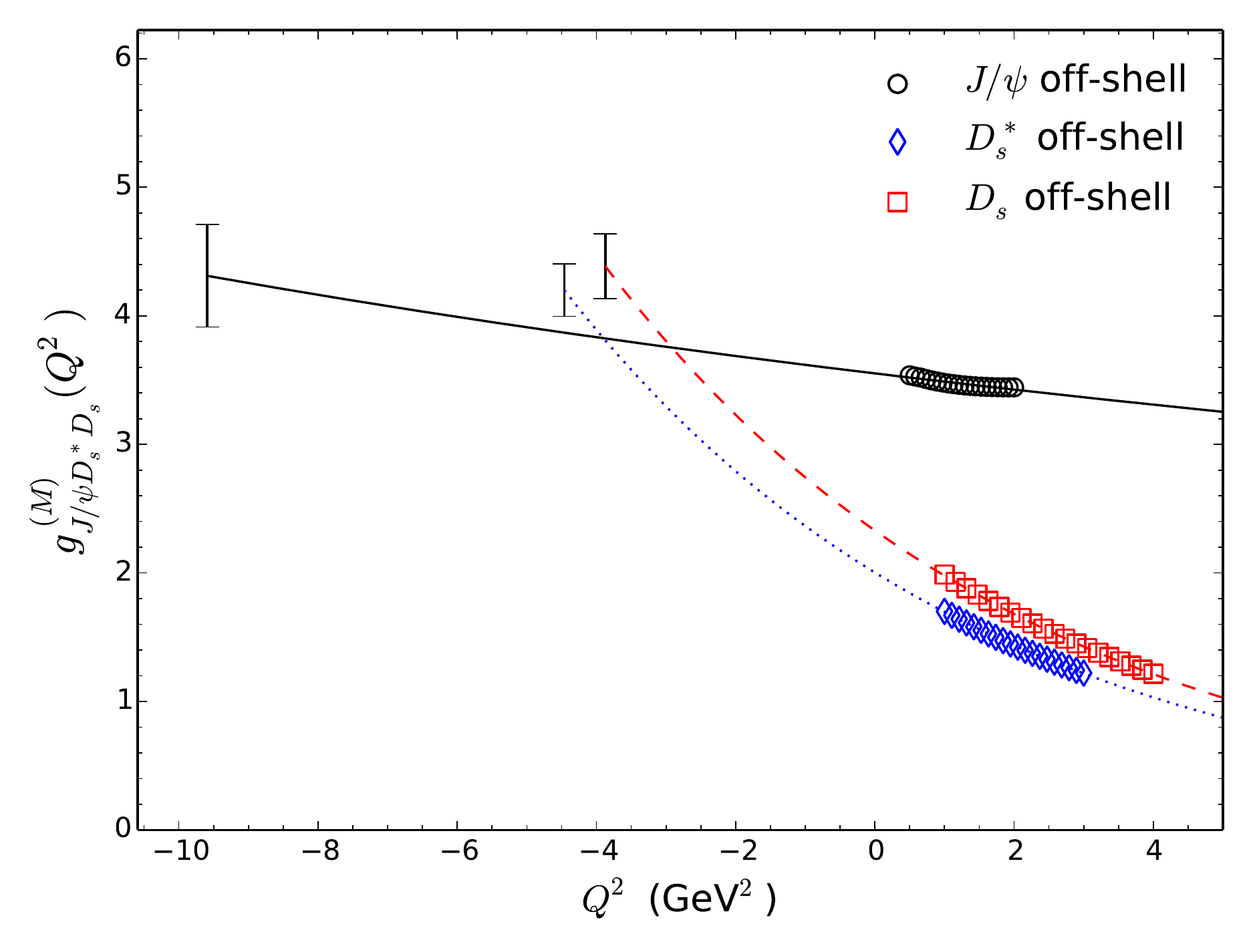}
  \caption{\label{fig:formfactors}Form factors of the $J/\psi D^*_s D_s$ vertex.}
  \end{figure}

The form factor obtained for the $J/\psi$ off-shell case was well adjusted by a monopolar curve, while for the $D^*_s$ and $D_s$ off-shell cases, the form factors were well adjusted by exponential curves (fig.~\ref{fig:formfactors}).

The coupling constants of each off-shell case are shown in table~\ref{tab:results}. They present different values among them, however, when the uncertainties are taken into account, they are all compatible, as suggested by fig.~\ref{fig:formfactors}, where the error bars stand for the uncertainties ($\sigma$) of the coupling constants also presented in table~\ref{tab:results}.

The uncertainties were calculated following the same procedures used in our previous works~\cite{Rodrigues2011127,CerqueiraJr2012130,OsorioRodrigues2014143}. They are estimated by studying the behavior of the coupling constants when each parameter is varied individually between its upper and lower limits. Experimental parameters have their own errors (shown in table~\ref{tab:errors}), while for the QCDSR parameters, we varied the thresholds $\Delta_s$ and $\Delta_u$ in $\pm 0.1$ GeV and the $Q^2$ window in $\pm 20\%$. For the error due to the Borel mass ($M^2$), we used the standard deviation of the form factor within the Borel window. After that, we calculated the average of all these contributions and their respective standard deviations to obtain the coupling constant and its uncertainty for a given off-shell case. In table~\ref{tab:errors}, we can see how the variance of each parameter affects the final value of the coupling constant.
\begin{table}[ht]
\begin{center}
\caption{\label{tab:errors}Percentage deviation of the coupling constants related to each parameter.}
\begin{tabular}{cccc}
\hline
\hline
 Parameter                    & $\Delta g_{J/\psi D^*_s D_s}^{(J/\psi)}$(\%) & $\Delta g_{J/\psi D^*_s D_s}^{(D^*_s)}$(\%) & $\Delta g_{J/\psi D^*_s D_s}^{(D_s)}$(\%)\\ 
\hline
$f_{D_s} = 257.5 \pm 6.1$(MeV)\cite{PDG}     & 1.94  & 1.93 & 1.94   \\
$f_{D^*_s} = 301 \pm 13$(MeV)$^a$     & 3.53  & 3.53 & 3.53   \\
$f_{J/\psi} = 416 \pm 6 $(MeV)\cite{PDG}     & 1.17  & 1.18 & 1.17 \\
$m_c = 1.27^{+0.07}_{-0.09}$ (GeV)\cite{PDG} & 19.5 & 12.9 & 16.6 \\
$m_s = 101^{+29}_{-21}$ (MeV)\cite{PDG}      & 2.56  & 0.96 & 2.25 \\ 
$M^2$ (GeV$^2$)$^b$     & 1.42 & 4.46 & 0.12 \\
$\Delta_s \pm 0.1$(GeV),$\Delta_u \pm 0.1$(GeV)    & 8.93  & 2.49 & 2.08  \\
$Q^2 \pm 20\%$ (GeV$^2$)$^b$            & 4.94 & 1.10 & 0.98  \\
$\langle \bar{s}s\rangle = -(290 \pm 15)^3$ (MeV$^3$)\cite{PhysRevD.87.034503} & 1.83 & - & -\\
$\langle g^2 G^2 \rangle = 0.88\pm 0.16$(GeV$^4$)\cite{Narison2012412} & 1.45 & 0.18 & 0.73\\
$\langle \bar{s}g\sigma \cdot G s \rangle = (0.8\pm 0.2)\langle \bar{s}s\rangle$(GeV$^5$)\cite{Ioffe2006232} & 0.49 & - & -\\
Fitting parameters  ($A$ and $B$)      & 14.9  & 0.12 & 0.06 \\
\hline
\hline
\end{tabular}
\end{center}
\hspace*{-5cm}$^a$Value obtained using $f_{D^*_s}/f_{D_s} = 1.17$ from \cite{PhysRevD.75.116001,PhysRevD.60.074501}.\\
\hspace*{-4.6cm}$^b$The intervals for these quantities are those of table~\ref{tab:results}.
\end{table}

According to table~\ref{tab:errors}, the variation of most of the parameters (the decay constants, condensates and the Borel mass for example) has a small impact on the value of the coupling constant. On the other hand, the uncertainty in the quark charm mass is the one with the biggest propagation over to the final value of the coupling constant. The fact that the sensitivity related to the Borel mass is small  was already expected, as we have very stable Borel windows (fig.~\ref{fig:estabilidade}). 

\section{Conclusion}

In this work we have calculated the $g_{J/\psi D^*_s D_s}$ coupling constant by three different QCD sum rules: one with the $D_s$ meson off-shell, another with the $D_s^*$ meson off-shell and a third one with the $J/\psi$  meson off-shell. This procedure allowed us to reduce the uncertainties related to the method, leading to compatible coupling constants, as seen in fig.~\ref{fig:formfactors}.

Taking the mean value between the numbers presented in table~\ref{tab:results}, we obtain the following final result for $g_{J/\psi D^*_s D_s}$:
\begin{align}
g_{J/\psi D^*_s D_s} = 4.30^{+0.42}_{-0.37}  \;\mbox{GeV$^{-1}$}\;.
\label{eq:ctejpsidsestds}
\end{align}

This coupling constant was obtained from sum rules that respect the criteria of the pole dominance over the continuum, the perturbative contribution of the OPE being the dominant one and the form factor stability regarding the Borel mass in the whole Borel window, as shown in figs.~\ref{fig:estabilidade} and~\ref{fig:jpsidsestdspolocont}.

Regarding the form factors, we observe some similarities with our previous work for the $J/\psi D_s D_s$ vertex~\cite{OsorioRodrigues2014143}. Among such similarities, there is the form factor given by a monopolar parametrization when the heaviest meson ($J/\psi$) is off-shell, while an exponential one is the case when one of the lightest mesons ($D_s$ or $D_s^*$) is off-shell. Furthermore, the OPE series presents, in both vertices, the same hierarchy for the contribution of each term, as well as comparable contributions among terms of the same dimension.

We can also compare our result for the $g_{J/\psi D^*_s D_s}$ coupling constant (eq.~(\ref{eq:ctejpsidsestds})) with the results of previous QCDSR works, presented in table~\ref{tab:otherresults}. 

\begin{table}[ht]
\caption{\label{tab:otherresults}Values of coupling constants obtained using different methods.}
\begin{tabular}{crl}
\toprule
Method and reference & \multicolumn{2}{c}{Coupling constant}\\
\midrule
 QCDSR~\cite{PhysRevD.89.016001}$\;$ & $g_{J/\psi D_s^* D_s} =$& $3.03\pm 0.62 \text{GeV}^{-1}$\\
 QCDSR~\cite{doi:10.1142/S0218301305003399}$\;\;\;$ & $g_{J/\psi D^* D}\;\; =$&  $4.0 \pm 0.6 \text{GeV}^{-1}$\\
 QCDSR~\cite{OsorioRodrigues2014143}$\;\;\;$ & $g_{J/\psi D_s D_s}\; =$& $5.98^{+0.67}_{-0.58}$\\
 VMD~\cite{PhysRevC.63.034901,PhysRevC.75.064903} & $g_{J/\psi D D}\;\;\;\; =$& $7.44$\\
\bottomrule
\end{tabular}
\end{table}

The comparison with ref.~\cite{PhysRevD.89.016001} is straightforward, its result differs from ours by $\approx 30\%$, both being compatible within $2\sigma$. Similar conclusions can be obtained from heavy quark effective theory (HQET)~\cite{PhysRevD.74.074003}, from where the relation $g_{J/\psi D^*_s D_s} = g_{J/\psi D_s D_s}/m_{D_s}$ stands, which allows us to compare the value of eq.~(\ref{eq:ctejpsidsestds}) with the result of our previous work~\cite{OsorioRodrigues2014143}. Using $m_{D_s} = 1.968$ GeV, we have $g_{J/\psi D_s D_s}/m_{D_s} = 3.04$ GeV$^{-1}$, a different result ($\approx 30\%$) from the $g_{J/\psi D^*_s D_s}$ in this work, but in agreement with the also sizable difference ($\approx 23\%$) found using the relation $g_{J/\psi D^* D} = g_{J/\psi D D}/m_{D}$ in~\cite{doi:10.1142/S0218301305003399}. A third comparison can be made invoking the SU(3) symmetry~\cite{PhysRevD.74.074003}: $g_{J/\psi D^*_s D_s} = g_{J/\psi D^* D}$, from which we can compare the result of eq.~(\ref{eq:ctejpsidsestds}) with the $g_{J/\psi D^* D}$ of ref.~\cite{doi:10.1142/S0218301305003399}. From this symmetry, we can see that these two coupling constants are compatible with each other within $1\sigma$, with a difference between them of $\approx 7\%$. This difference is comparable with the one found using the SU(3) relation $g_{J/\psi D_s D_s} = g_{J/\psi D D}$ in ref.~\cite{OsorioRodrigues2014143}. 

Finally, we can compare our result with the coupling constant $g_{J/\psi D D}$ from the vector meson dominance (VMD) \cite{PhysRevC.63.034901,PhysRevC.75.064903}  using the SU(3) and HQET relations, which leads to $g_{J/\psi D^*_s D_s} = 3.84$ GeV$^{-1}$. This coupling is also compatible with the one of eq.~(\ref{eq:ctejpsidsestds}) within $2\sigma$.

\section*{Acknowledgements}
This work has been partially supported by CNPq and CAPES.

\bibliography{bibliografia}

\appendix
\section{Full expressions for $\langle g^2 G^2\rangle$ and $\langle \bar{q}g\sigma G q \rangle$ correlators - $J/\psi$ off-shell case.}
\label{appendix:remaningexpressions}

Equation~(\ref{eq:JpsiDsEstDs-Jpsioff-GG}) is the double Borel transform of the gluon condensate contribution to the OPE.
\begin{multline}
\mathcal{B}\mathcal{B} \left[\Gamma^{\left < g^2G^2 \right >}\right ] = \sum_{x=d,e,f,g,h,i} \mathcal{B}\mathcal{B} \left[\Gamma^{\left < g^2G^2 \right >}_{(x)}\right] = \\
-\frac{\left < g^2G^2 \right >}{96\pi^2} 
\int_{\frac{1}{M'^2}}^\infty d{\alpha_1} e^{\frac{\alpha_1 M'^2t-t}{M^2+M'^2} + \frac{\left (\alpha_1 +\alpha_1\frac{M'^2}{M^2} \right )\left [ m_c^2- m_s^2 - \alpha_1^2m_c^2M'^2\right]}{\alpha_1M'^2-1}  }
p'^\lambda p^\sigma \varepsilon_{\mu\nu\lambda\sigma} \sum_{x=d,e,f,g,h,i} F_{(x)}(\alpha_1),
\label{eq:JpsiDsEstDs-Jpsioff-GG}
\end{multline}
where the index $x = d\text{ to } i$  corresponds to the diagrams of fig.~\ref{subfig:d}-\ref{subfig:i}, with $F_{(d)}\text{-}F_{(i)}$ given by eqs.~(\ref{eq:Fd})-(\ref{eq:Fi}) respectively.
\begin{align}
F_{(d)} = m_c(2 \alpha_1^2 m_c^2 M^2-3 \alpha_1 M^2+2 \alpha_1 m_c m_s- 2 \alpha_1 m_c^2+3)/4
\label{eq:Fd}
\end{align}
\begin{multline}
F_{(e)} = m_c(\alpha_1^3 t M^4-2 \alpha_1^2 M^4-2 \alpha_1^2 t M^2+4 \alpha_1^2 m_c m_s M^2+\\6 \alpha_1 M^2+\alpha_1 t+4 \alpha_1 m_s^2 -4 \alpha_1 m_c m_s -4 )/(8(\alpha_1 M^2-1))
\end{multline}
\begin{align}
F_{(f)} = (\alpha_1 m_c M^2+m_s-m_c)/2
\end{align}
\begin{align}
F_{(g)} = 2m_s (2 \alpha_1^2 m_c m_s M^2-3 \alpha_1 M^2+2 \alpha_1 m_s^2-2 \alpha_1 m_c m_s +3)/(\alpha_1 M^2-1)^3
\end{align}
\begin{align}
F_{(h)} = -(2 \alpha_1 m_c M^2+3 m_s-2 m_c)/(\alpha_1 M^2-1)
\end{align}
\begin{align}
F_{(i)} = (\alpha_1 m_c M^2+m_s-m_c)/(\alpha_1 M^2-1)
\label{eq:Fi}
\end{align}

The double Borel transforms of the mixed quark-gluon condensate contributions are given by eqs.~(\ref{eq:JpsiDsEstDsJpsioffqGq}) and~(\ref{eq:pijpsidsestdsJpsiOffmqGq}).
\begin{multline}
\mathcal{B}_{M^2} \mathcal{B}_{M'^2}\left [ \Gamma^{\left<\bar{s} g\sigma G s \right >} \right ] = 
 -\frac{\left < \bar{s}g\sigma G s \right >}{12M'^4M^4} \left ( 3\left [Q^2M'^2M^2 
 + m_c^2M^4 + 2m_c^2M'^2M^2+m_c^2M'^4 \right ] - M^2M'^4 \right.\\ 
\left.  + 3M^4M'^2 \right ) p'^\lambda p^\sigma \varepsilon_{\mu\nu\lambda\sigma} e^{-\frac{m_c^2}{M^2}} e^{-\frac{m_c^2}{M'^2}}.
\label{eq:JpsiDsEstDsJpsioffqGq}
\end{multline}

\begin{multline}
\mathcal{B}_{M^2} \mathcal{B}_{M'^2}\left [\Gamma^{m_{s} \left<\bar{s} g\sigma G s \right >} \right ] =
 -\frac{m_{s} \left<\bar{s} g\sigma G s \right >}{24M^6M'^6} \left ( 2m_c^3\left [ M^6 + 3M^4M'^2   
 + 3M^2M'^4 + M'^6\right ] + \right. \\ \left. m_cM^2M'^2 \left [ 12M^4 + 6M^2M'^2 + M'^4\right ]\right ) 
p'^\lambda p^\sigma \varepsilon_{\mu\nu\lambda\sigma} e^{-\frac{m_c^2}{M^2}} e^{-\frac{m_c^2}{M'^2}}.
\label{eq:pijpsidsestdsJpsiOffmqGq}
\end{multline}

\end{document}